\pdfoutput=1 
\documentclass[%
aps,
superscriptaddress,
bibnotes,
amsmath,amssymb,
prr,
floatfix,
longbibliography,
twocolumn,
]{revtex4-2}

\usepackage{float}
\usepackage{color}
\usepackage[usenames,dvipsnames,svgnames,table]{xcolor}
\usepackage[colorlinks=true,linkcolor=blue,urlcolor=blue,citecolor=blue]{hyperref}
\usepackage{graphicx}
\usepackage{dcolumn}
\usepackage{bm}
\usepackage{xcolor}
\usepackage{enumerate}
\usepackage{braket}
\usepackage[T1]{fontenc}
\usepackage{listings}
\usepackage{tabularx}
\usepackage{booktabs}
\usepackage{amsmath}
\usepackage[caption=false]{subfig}
\usepackage{ragged2e} 
\DeclareCaptionJustification{justified}{\justifying}

\usepackage[english]{babel}
\makeatletter
\def\bbl@set@language#1{%
  \edef\languagename{%
    \ifnum\escapechar=\expandafter`\string#1\@empty
    \else\string#1\@empty\fi}%
  \@ifundefined{babel@language@alias@\languagename}{}{%
    \edef\languagename{\@nameuse{babel@language@alias@\languagename}}%
  }%
  \select@language{\languagename}%
  \expandafter\ifx\csname date\languagename\endcsname\relax\else
    \if@filesw
      \protected@write\@auxout{}{\string\select@language{\languagename}}%
      \bbl@for\bbl@tempa\BabelContentsFiles{%
        \addtocontents{\bbl@tempa}{\xstring\select@language{\languagename}}}%
      \bbl@usehooks{write}{}%
    \fi
  \fi}
\newcommand{\DeclareLanguageAlias}[2]{%
  \global\@namedef{babel@language@alias@#1}{#2}%
}
\makeatother

\DeclareLanguageAlias{en}{english}

\newcommand{\br}{\bm{r}}
\newcommand{\bR}{\bm{R}}
\newcommand{\ubr}{\underline{\bm{r}}}
\newcommand{\ubR}{\underline{\bm{R}}}
\newcommand{\ubp}{\underline{\bm{p}}}
\newcommand{\bp}{{\bm{p}}}

\newcommand{\s}{_\mathrm{{\scriptscriptstyle S}}}

\newcommand{\xc}{_\mathrm{{\scriptscriptstyle XC}}}
\newcommand{\ee}{_{{\scriptscriptstyle ee}}}
\newcommand{\ei}{_{{\scriptscriptstyle ei}}}
\newcommand{\ii}{_{{\scriptscriptstyle ii}}}
\newcommand{\Ne}{{N_{{\scriptscriptstyle e}}}}
\newcommand{\Ni}{{N_{{\scriptscriptstyle i}}}}

\newcommand{\kB}{k_\mathrm{B}}

\begin{document}

\title{Accelerating Equilibration in First-Principles Molecular Dynamics with Orbital-Free Density Functional Theory}

\author{Lenz Fiedler}
\email{l.fiedler@hzdr.de}
\affiliation{Center for Advanced Systems Understanding (CASUS), D-02826 G\"orlitz, Germany}
\affiliation{Helmholtz-Zentrum Dresden-Rossendorf, D-01328 Dresden, Germany}

\author{Zhandos A. Moldabekov}
\email{z.moldabekov@hzdr.de}
\affiliation{Center for Advanced Systems Understanding (CASUS), D-02826 G\"orlitz, Germany}
\affiliation{Helmholtz-Zentrum Dresden-Rossendorf, D-01328 Dresden, Germany}

\author{Xuecheng Shao}
\email{xuecheng.shao@rutgers.edu}
\affiliation{Department of Chemistry, Rutgers University, 73 Warren St., Newark, NJ 07102, USA}

\author{Kaili Jiang}
\email{kaili.jiang@rutgers.edu}
\affiliation{Department of Chemistry, Rutgers University, 73 Warren St., Newark, NJ 07102, USA}

\author{Tobias Dornheim}
\email{t.dornheim@hzdr.de}
\affiliation{Center for Advanced Systems Understanding (CASUS), D-02826 G\"orlitz, Germany}
\affiliation{Helmholtz-Zentrum Dresden-Rossendorf, D-01328 Dresden, Germany}

\author{Michele Pavanello}
\email{m.pavanello@rutgers.edu}
\affiliation{Department of Chemistry, Rutgers University, 73 Warren St., Newark, NJ 07102, USA}
\affiliation{Department of Physics, Rutgers University, 101 Warren St., Newark, NJ 07102, USA}

\author{Attila Cangi}
\email{a.cangi@hzdr.de}
\affiliation{Center for Advanced Systems Understanding (CASUS), D-02826 G\"orlitz, Germany}
\affiliation{Helmholtz-Zentrum Dresden-Rossendorf, D-01328 Dresden, Germany}

\date{\today}

\begin{abstract}
We introduce a practical hybrid approach that combines orbital-free density functional theory (DFT) with Kohn-Sham DFT for speeding up first-principles molecular dynamics simulations. Equilibrated ionic configurations are generated using orbital-free DFT for subsequent Kohn-Sham DFT molecular dynamics. This leads to a massive reduction of the simulation time without any sacrifice in accuracy. We assess this finding across systems of different sizes and temperature, up to the warm dense matter regime. To that end, we use the cosine distance between the time series of radial distribution functions representing the ionic configurations.
Likewise, we show that the equilibrated ionic configurations from this hybrid approach significantly enhance the accuracy of machine-learning models that replace Kohn-Sham DFT. 
Our hybrid scheme enables systematic first-principles simulations of warm dense matter that are otherwise hampered by the large numbers of atoms and the prevalent high temperatures. Moreover, our finding  provides an additional motivation for developing  kinetic and noninteracting free  energy functionals for orbital-free DFT.   
\end{abstract}

\maketitle


\section{Introduction}
\label{sec:introduction}
After decades of advances and methodological improvements, density functional theory (DFT)~\cite{hohenberg_inhomogeneous_1964} has become the computational tool of choice for solving the majority of computational materials science problems from first principles. It delivers impactful predictions and insights that propel scientific progress due to its balance of computational cost and accuracy~\cite{DFTReview}.  
Yet, new scientific problems challenge even the most efficient DFT implementations. Over the last decades, understanding materials under extreme conditions, i.e., at high temperatures and large pressures, has become an emerging field of research.  Most notably, studying \textit{warm dense matter} (WDM) is currently in the spotlight~\cite{POP2020, WDM1,WDM2}. In WDM both the Wigner-Seitz radius $r_s$ and the reduced temperature $\theta=\tau/\tau_\mathrm{F}$ are close to unity, where $\tau_\mathrm{F}$ denotes the Fermi temperature. This poses a challenge, because Coulomb correlations, fermionic exchange, and thermal excitations are all equally relevant. While modeling WDM using classical approaches is thus inaccurate, the use of quantum mechanical methods, such as DFT, is computationally expensive. 

\begin{figure}[tp]
    \centering
    \includegraphics[width=0.9\columnwidth]{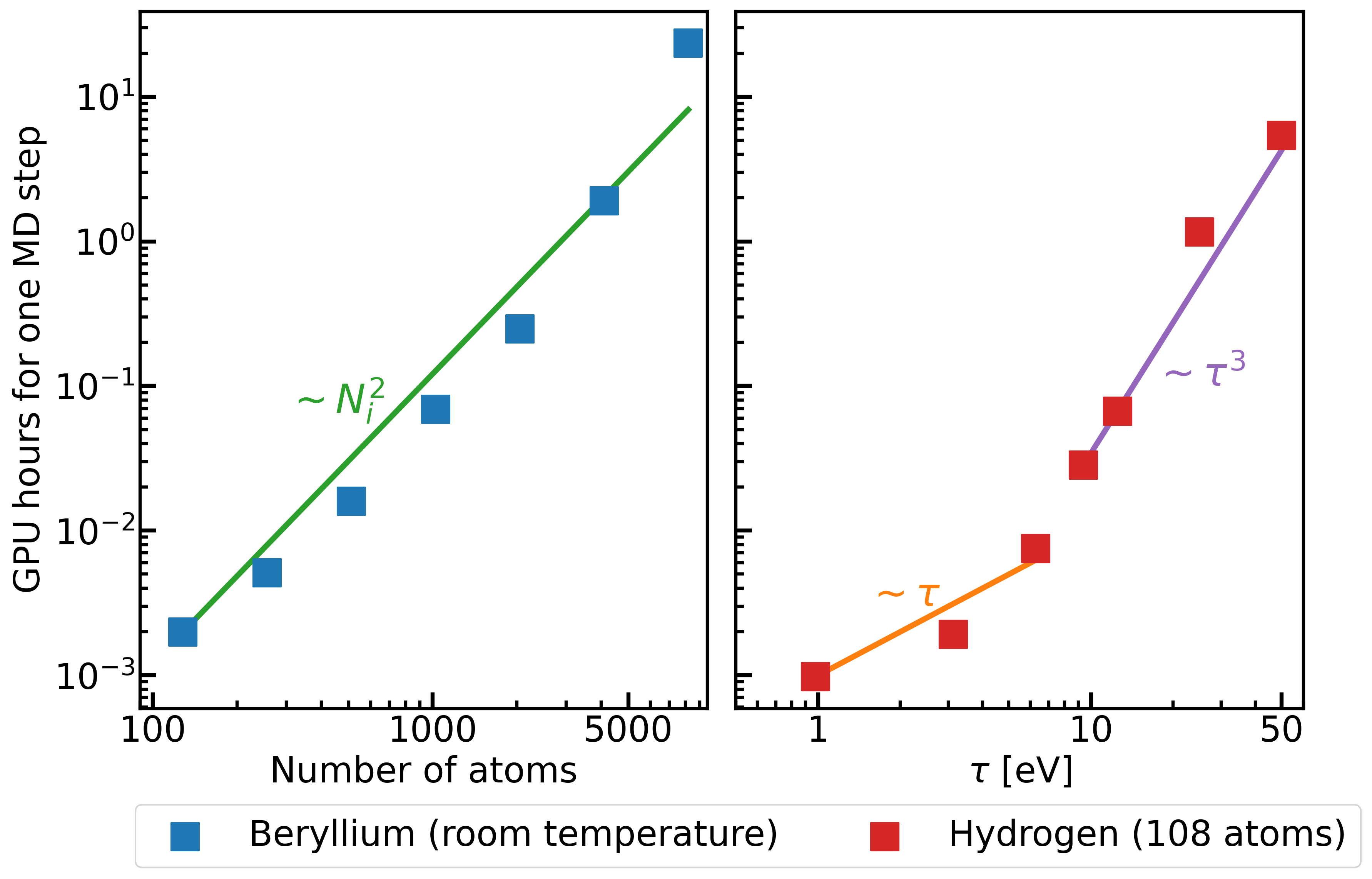}
    \caption{The scaling behavior of KS-DFT as a function of system size for Beryllium at room temperature (left) and as a function of temperature for Hydrogen (right). The computational parameters of the KS-DFT-MD calculations are provided in Sec.~\ref{sec:compdetails}. Please note that the apparent decrease in performance when going from 4096 to 8192 atoms is caused by the necessity to move from one compute node to two for simulations of this size. Thus, the simulation of 8192 Beryllium atoms is the only one affected by an additional communication penalty.}
    \label{fig:motivation_temp}
\end{figure}

Besides theoretical hurdles in applying Kohn-Sham DFT (KS-DFT)~\cite{kohn_self-consistent_1965} to matter under extreme conditions, one is quickly confronted with seemingly insurmountable scaling limitations. KS-DFT scales unfavorably with growing temperatures $\tau$, as illustrated in Fig.~\ref{fig:motivation_temp}. For small temperatures, we observe a linear growth of computation time with temperature, most likely due to efficient algorithms used in the employed DFT code. As we move to larger temperatures, the expected cubic scaling of DFT with growing temperature can be observed, which makes simulations in the WDM regime, i.e., at $\theta \gtrapprox 1$ often impractical. Approaches to circumvent this shortcoming and achieve DFT calculations that scale linearly with temperature are an area of active research~\cite{lineartemp1, lineartemp2, lineartemp3}.

Likewise modeling WDM requires extended length scales, i.e., large simulation cells to alleviate potential finite-size problems. While there exist approaches that permit DFT calculations to scale linearly with system size~\cite{nakata_large_2020,hernandez_self_1995,vandevondele_linear_2012,mohr2015accurate,galli_electronic_1994}, usually based on density matrix approaches~\cite{density_matrix_paper}, standard DFT codes are capable of retaining only a quadratic scaling behavior with respect to number of particles (see Fig.~\ref{fig:motivation_temp}). This quadratic scaling becomes intractable quickly, when pushing calculations towards thousands of atoms.

Overcoming these limitations has been subject of active research. One option is orbital-free DFT (OF-DFT)~\cite{ligneres_introduction_2005}, which scales more favorably than KS-DFT both with temperature and system size~\cite{DFTpy}, but suffers from reduced accuracy due to approximating the kinetic energy functional. The quest for increasingly accurate kinetic energy functionals is ongoing~\cite{PhysRevB.75.155109, doi:10.1021/ct400836s, PhysRevB.100.125106, PhysRevB.100.125107}. Due to these developments, structural properties of certain elements of interest for WDM can now be computed rather accurately with orbital-free DFT molecular dynamics (OF-DFT-MD) both at ambient~\cite{DFTpy} and extreme conditions~\cite{karasiev_nonempirical_2013}. 

Another emerging research area is the use of machine learning (ML) models~\cite{deepdive} in electronic structure theory. They are capable of massively reducing the computational cost for computing properties at KS-DFT accuracy, but can often suffer from generalization errors. In order to achieve high fidelity, these models need to be initialized with ionic configurations close to those trained on.  

When standard KS-DFT molecular dynamics (KS-DFT-MD) simulations are carried out at finite temperatures, a significant amount of compute time goes into the equilibration of the ionic configuration, creating an unnecessary computational overhead.
We show that this overhead can be avoided by optimizing the initial ionic configuration. This can be achieved in practice in terms of a hybrid approach that combines OF-DFT-MD with KS-DFT-MD. By using OF-DFT-MD to carefully initialize KS-DFT-MD trajectories, we significantly reduce the computational overhead, both for large simulation cells and high temperatures.

The paper is organized as follows. In Section~\ref{sec:theoretical_background}, we provide the theoretical background and describe our computational and analysis methods. In Section~\ref{sec:results}, we present the central results of our hybrid approach. These include initializing MD trajectories for ML methods, achieving rapid equilibration in MD simulations with large atom counts, and enabling efficient KS-DFT-MD simulations at high temperatures. We have chosen three different systems to represent these fields of application. For the application to ML methods we show results for Aluminum at room temperature (drawing on findings presented in Ref.~\cite{ellis_accelerating_2021}), for the simulation of extended length scales we chose Beryllium at room temperature, while for the treatment of large temperatures we consider Hydrogen, which is a highly relevant material in the WDM regime. In Section~\ref{sec:end}, we summarize our findings and conclude with an outlook on future research. 

\section{Methods}
\label{sec:theoretical_background}
\subsection{Density Functional Theory}
In this work we deal with two flavors of DFT, namely KS-DFT and OF-DFT. Both are computational methods for treating a system of $\Ni$ ions at collective positions $\ubR$ and $\Ne$ electrons at collective positions $\ubr$, governed by a many-body Schr\"odinger equation
\begin{equation}
	\hat{H}(\ubr; \ubR) \Psi(\ubr; \ubR) = E \Psi(\ubr; \ubR) \label{eq:Schrödinger_stationary}
\end{equation}
with the Hamiltonian
\begin{align}
    \hat{H}(\ubr; \ubR) = &\hat{T}_{{\scriptscriptstyle e}}(\ubr) + \hat{T}_{{\scriptscriptstyle i}}(\ubR) \nonumber \\ &+ \hat{V}\ee(\ubr) + \hat{V}\ei(\ubr; \ubR)+ \hat{V}\ii(\ubR) \label{eq:fullHamiltonian}
\end{align}
and the electron-ion wave function $\Psi(\ubr; \ubR)$. In Eq.~(\ref{eq:fullHamiltonian}), $\hat{T}_\mathrm{{\scriptscriptstyle e}}$ and $\hat{T}_\mathrm{{\scriptscriptstyle i}}$ denote the kinetic energy of electrons and ions, respectively, while $\hat{V}\ee(\ubr)$, $\hat{V}\ei(\ubr; \ubR)$ and $\hat{V}\ii(\ubR)$ denote the electron-electron, electron-ion and ion-ion interaction, respectively. The nature of these interactions leads to Eq.~(\ref{eq:Schrödinger_stationary}) being computationally intractable. In both KS-DFT and OF-DFT, two important concepts are employed to make the coupled electron-ion problem manageable. Firstly, the Born-Oppenheimer approximation~\cite{born_zur_1927} is employed, which separates Eq.~(\ref{eq:Schrödinger_stationary}) into an ionic and an electronic problem. This is feasible since the ions are much heavier than the electrons, resulting in a much larger time scale for ionic motion compared to electronic. The ions are therefore treated as classical point particles, see Sec.~\ref{sec:molecular_dynamics}. This leads to a  Born-Oppenheimer Hamiltonian for the electronic problem,
\begin{align}
\label{eq.bo-hamiltonian}
\hat{H}^{BO}(\ubr; \ubR) = &\hat{T}^{e}(\ubr) + \hat{V}^{ee}(\ubr) \nonumber \\ &+ \hat{V}^{ei}(\ubr; \ubR)+ E^{ii}(\ubR)
\end{align}
that now depends only parametrically on $\ubR$. 
Secondly, both flavors of DFT rely on the Hohenberg-Kohn theorems~\cite{hohenberg_inhomogeneous_1964} which provide a one-to-one correspondence between the ground-state electronic density $n_0(\br)$ and the external potential generated by the ions $V\ei(\br)$. Therefore, all properties of the system defined by the Born-Oppenheimer Hamiltonian can be determined by knowledge of the ground state density. 

The total energy functional in both OF-DFT and KS-DFT is expressed as
\begin{equation}
E_\mathrm{tot}[n] = T\s[n] + U[n] + E\xc[n] + V\ei[n] + E\ii \label{eq:Etot.zerotemp}\; ,
\end{equation}
where $T\s$ denotes the single-particle kinetic energy, $U$ the Hartree energy, i.e., the electrostatic interaction of the density with itself, $E\xc$ the exchange-correlation energy that captures energetic contributions of the electron-electron interaction not included in $U$, $V\ei[n]$ the electron-ion interaction energy, and $E\ii$ the ion-ion interaction energy, which amounts to a constant shift in energy. 

In practice, $E\xc[n]$ has to be approximated, and the accuracy of DFT calculations depends primarily on the choice of approximation. A plethora of suitable functionals exists. They either depend solely on the electronic density such as the local density approximation (LDA)~\cite{kohn_self-consistent_1965,ceperley_ground_1980}), incorporate further quantities such as the density gradient in the generalized gradient approximation (GGA) functionals~\cite{perdew_accurate_1986,perdew_accurate_1992,perdew_generalized_1996}, or also depend on the kinetic energy density in meta-GGAs such as the SCAN functional~\cite{sun_strongly_2015}.

The challenge in DFT is to identify $n_0$. This is achieved by minimizing the total energy functional with respect to variations in the density $n$. However, a central problem is the lack of an exact expression for $T\s[n]$ as an explicit density functional, although such an expression does exist in terms of a potential functional for the density~\cite{PhysRevLett.106.236404}. 
The manner in which $T\s[n]$ is expressed, is the key difference between OF-DFT and KS-DFT. In KS-DFT, $T\s[n]$ is not approximated. Instead the minimization of the total energy functional is performed with respect to single-particle orbitals $\phi_j(\br)$ yielding the KS equations~\cite{kohn_self-consistent_1965}
\begin{equation}
\left[-\frac{1}{2}\nabla^2 + v\s(\br)\right]\phi_j(\br) = \epsilon_j \phi_j(\br)\,, \label{eq:KSequation.zerotemp}
\end{equation}
where $\epsilon_j$ denote the single-particle eigenvalues and $v\s$ the Kohn-Sham potential. This auxiliary system is restricted to reproduce the density of the interacting system via 
\begin{align}
\label{eq:density.dft0K}
n(\br) &= \sum_{j=0}^{\Ne} \, f_j |\phi_j(\br)|^2 \; ,
\end{align}
with the occupation numbers $f_j$. By determining the KS potential $v\s$ self-consistently, it is ensured that Eq.~(\ref{eq:density.dft0K}) recovers the density of the interacting system. 
The KS kinetic energy is determined exactly in terms of the KS orbitals $\phi_j(\br)$  as 
\begin{equation}
T\s =\sum_{j=0}^\Ne \int d\br\, \phi_j^*(\br) \left(-\frac{1}{2}\nabla^2\right) \phi_j(\br) \; .
\end{equation}

OF-DFT~\cite{ligneres_introduction_2005,wesolowski_recent_2012,karasiev_progress_2014} follows a different route by approximating the kinetic energy functional \textit{directly} as a functional of the density. Similar to $E\xc[n]$, various approximations do exist and are being developed. Commonly used kinetic energy functionals are built using a combination of the Thomas-Fermi functional $T^\mathrm{TF}$~\cite{thomas_1927,fermi1928statistische}, the von Weizsäcker functional $T^\mathrm{vW}$~\cite{weizsacker1935theorie}, and a non-local term $T^\mathrm{NL}$, i.e.,
\begin{equation}
T\s[n] \approx T_\mathrm{TF}[n] + T_\mathrm{vW}[n] + T_\mathrm{NL}[n] \; , \label{eq:kinetic_energy_functional_ofdft}
\end{equation}
with
\begin{align}
T_\mathrm{TF}[n] &=  \frac{3}{10}{(3\pi^2)}^{2/3}\int \,d\br\, n^{5/3}(\br) \; , \label{eq:tf_ke}\\ 
T_\mathrm{vW}[n] &= \frac{1}{2} \int \,d\br\, \nabla n^{1/2}(\br) \cdot {\nabla n}^{1/2}(\br)\;.
\label{eq:vW}
\end{align}
A number of approximations exists for $T^\mathrm{NL}$, e.g., the Wang-Teter~\cite{wang-teter}, Mi-Genova-Pavanello~\cite{mi-genova-pavanello}, and Wang-Govind-Carter~\cite{wang-govind-carter} functionals. 

The difference in how the kinetic energy is treated results in contrasting differences for KS-DFT and OF-DFT in terms of accuracy and computational cost. KS-DFT scales formally as $\sim \Ne^3$, although in practice efficient codes scale with $\sim \Ne^2$, due to need to solve the KS equations self-consistently. On the other hand, OF-DFT scales linearly with $\Ne$. Generally, KS-DFT outperforms OF-DFT in terms of accuracy. 

Finally, often temperature has to be taken into account in DFT calculations. Finite-temperature DFT for $\tau>0$K~\cite{mermin_thermal_1965, pittalis_exact_2011, karasiev_nonempirical_2013, graziani_thermal_2014} follows the equations outlined above. One significant change is that the role of the total energy in Eq.~(\ref{eq:Etot.zerotemp}) is replaced by the total free energy 
\begin{align}
A_\mathrm{tot}[n] =& T\s[n] - \kB\tau S\s[n] + U[n] \nonumber \\ &+ E^\tau\xc[n] + V\ei[n] + E\ii\; , \label{eq:Etot.nonzerotemp}
\end{align}
which includes the single-particle entropy $S\s[n]$. In finite-temperature DFT calculations, one seeks to find $n$ such that $A_\mathrm{total}$ is minimal.
OF-DFT and KS-DFT vary in their evaluation of $S\s$. In OF-DFT, an explicit density functional is employed to evaluate this term. It is constructed following the same principles as for $T\s[n]$~\cite{OFDFTentropy2, OFDFTentropy1}, while in KS-DFT this term is evaluated exactly as
\begin{align}
S\s = &- \sum_{j=0}^\Ne \big[ f^\tau(\epsilon^\tau_j) \ln{\left(f^\tau(\epsilon^\tau_j)\right)} \nonumber \\ &+ (1-f^\tau(\epsilon^\tau_j))\ln\left(1-f^\tau(\epsilon^\tau_j)\right) \big] \;  
\end{align}
with the now temperature-dependent occupation number $f^\tau(\epsilon)$ and the KS energy eigenvalues $\epsilon^\tau_j$. Also the XC energy becomes explicitly temperature dependent. That dependence is, however, often neglected.

The computational cost of KS-DFT increases significantly with increasing temperature, because a larger number of KS orbitals needs to be included in the evaluation of the density, which is done via Eq.~(\ref{eq:density.dft0K}), with the occupation numbers given by the Fermi-Dirac distribution $f^\tau(\epsilon_j^\tau)$. Temperature enters this evaluation both through the temperature dependence of $f^\tau$ as well the Kohn-Sham eigenvaues $\epsilon_j^\tau$. Contrarily, in OF-DFT the temperature has virtually no effect on the computational cost.

\subsection{Molecular Dynamics}
\label{sec:molecular_dynamics}
In most MD simulations, the ions are considered as classical point particles. In first-principles MD simulations, the electrons are treated on the quantum level using DFT. ML trained interatomic potentials~\cite{thompson_spectral_2015,schutt_schnet_2018,bartok_gaussian_2010} are also gaining traction. They encode the potential energy surface obtained from DFT in an efficient manner and, thus, enable accurate MD simulations at DFT accuracy at large scales. In each time-step $t$, each ion $\alpha=1...\Ni$ with mass $m_\alpha$ is time evolved on the potential energy surface $A_\mathrm{tot}$ in terms of the Newtonian equations of motion
\begin{equation}
m_\alpha \frac{d^2 \bR_\alpha}{d t^2} = -\frac{\partial A_\mathrm{tot}}{\partial \bR_\alpha} \; . \label{eq:NewtonianMotion}
\end{equation} 
The right-hand side of Eq.~(\ref{eq:NewtonianMotion}) represents the atomic forces that are determined from either OF-DFT or KS-DFT. After the ionic positions have been updated, a subsequent DFT calculation determines new atomic forces, and uses these to update the ionic positions; such a loop is continued until the desired number of time steps has been performed. 

In practice, such MD simulations are realized via thermostats, which ensure sampling according to a proper thermodynamic ensemble, by conserving certain physical quantities. Within this work, we perform DFT-MD simulations in the canonical ($NVT$) ensemble, i.e., the number of particles, volume of the simulation cell, and temperature are fixed. In practice, this is realized via the Nos\'{e}-Hoover algorithm~\cite{hoover,nose}. Here, one utilizes a heat bath coupled to the system to ensure constant particle number, volume, and temperature. The coupling is realized in terms of additional terms that are added to the (classical) Hamiltonian of the system, i.e., 
\begin{equation}
    H_\mathrm{NH}=H_0(\ubR,\ubp/s)+N_\mathrm{df}\kB\tau\ln{s}+\frac{{\bp_s}^2}{2Q} \; , \label{eq:NHHamiltonian}
\end{equation}
where $H_0$ is the Hamiltonian of the classical system of ions with collective momenta $\ubp$, $N_\mathrm{df}$ the number of degrees of freedom coupled to the thermostat, and $s$ an artificial variable that represents the heat bath with momentum $\bp_s$. Introducing $s$ in Eq.~(\ref{eq:NHHamiltonian}) implies a coordinate transformation, i.e., a scaling of the momenta in the system. This framework gives rise to updated equations of motions~\cite{evans1985nose, siboni2016molecular} for the $\alpha$-th ion
\begin{align}
    \dot{\bR} &= \frac{\bp_\alpha}{m_\alpha s^2} \; , \\
    \dot{\bp}_\alpha &= -\frac{\partial A_\mathrm{tot}}{\partial \bR_\alpha} \; ,\\
    \dot{s} &= \frac{\bp_s}{Q} \; ,\\
    \dot{\bp_s} &=\sum_{\alpha=1}^\Ni \left( \frac{{\bp_\alpha}^2}{m_\alpha s^3} \right) -\frac{N_\mathrm{df} \kB\tau}{s} \; ,
\end{align}
where the generalized coordinates $\bm{q}$ are set to $\ubR$ for the sake of simplicity. This system of equations shows that the strength of the coupling of the heat bath to the system is governed by $Q$, which represents an imaginary mass. $Q$ is the principal tuning parameter of the thermostat and has to be carefully chosen to ensure proper sampling. The larger $Q$, the less strongly coupled heat bath and simulated system become, and the DFT-MD simulation eventually generates a micro-canonical ensemble. Conversely, too small values of $Q$ introduce unphysical, periodical temperature oscillations~\cite{nhoscillations}.

\subsection{Equilibration analysis}
\label{sec:equianalysis}
We now explain how OF-DFT can be used to reduce the computational cost of KS-DFT-MD simulations. At the center of attention is the process of \textit{equilibration}, i.e., the period of a KS-DFT-MD trajectory during which a system is brought to the desired temperature. As we aim to specifically shorten this period of the simulation, we need a systematic approach to determine its length. Equilibration is often done heuristically, i.e., if a trajectory has been propagated for a sufficient time, or if configurations appear reasonably equilibrated, one can start sampling thermodynamic observables. Here, we need to make sure that we systemtatically capture the first time-step that can be considered equilibrated. We do so in a transferable fashion, as to make assertions on the usefulness of OF-DFT-MD trajectories for KS-DFT-MD simulations. We thus introduce an analysis based on the ionic radial distribution function (RDF), which is defined as the average number of ions $\Xi(r)$ contained in a hollow shell of radius $r+dr$ and volume $V_\mathrm{shell}$, for a cell with $\Ni$ ions, normalized by the ionic density $\rho=\Ni/V_\mathrm{cell}$~\cite{rdf1, rdf2, rdf3}, i.e.,
\begin{equation}
    g(r) = \frac{\Xi(r)}{\rho \Ni V_\mathrm{shell}} \; ,
\end{equation}
which can be shown to be equivalent to
\begin{equation}
g(r) = \frac{\Xi(r)}{4\rho\pi dr \left(r^2+\frac{{dr}^2}{12}\right)} \label{eq:rdf} \; .
\end{equation}
Generally, the RDF is useful to distinguish between different phases or structures. Here, we follow the assumption that if two structures have reasonably close RDFs, they belong to similar phase of a material, i.e., if one of them can be considered equilibrated, so can the other. Our analysis, thus, centers on calculating and comparing RDFs across the trajectory. This poses the problem of choosing an appropriate reference point. One possible approach is averaging the RDF over the end of the trajectory. However, this would necessitate the comparison of such an averaged RDF with noisier RDFs at the beginning of the trajectory, requiring additional averaging. 

We therefore follow a slightly different approach and assign the very last ionic configuration of the trajectory as the reference point. Thereafter, the cosine distance between the RDFs of all ionic configurations $g_a=g(r)[\ubR_a]$ (where $a$ denotes the number of the time-step) within this trajectory and the reference RDF $g_\mathrm{ref}=g(r)[\ubR_\mathrm{ref}]$ is calculated in terms of
\begin{equation}
    \tilde{d}^a_C\left(g_\mathrm{ref}, g_a\right) = 1-\frac{g_\mathrm{ref}\cdot g_a}{\|{g_\mathrm{ref}}\| \|{g_a}\|} \label{eq:cosine_distance} \; .
\end{equation}
Here, $g_\mathrm{ref}\cdot g_a$ is the dot product of two RDFs expressed as vectors with dimensionality $r_\mathrm{max}$, where $r$ is discretized as $r=0,\,dr,\,2dr,\,...,\,r_\mathrm{max}$. Cosine distances (or conversely, cosine similarities) of data points are a standard analysis technique in data science~\cite{cosine_distance_reference}. The cosine distance is a measure of how different two vectors are. It becomes 0 for identical vectors, and 1 for entirely dissimilar vectors.

The resulting signal $d^a_C$ is noisy, as raw RDFs are compared to one another. We thus smoothen $d^a_C$ using a running average of width $\sigma$. Each data point is reassigned as an arithmetic average of the $\sigma-1$ preceding data points and itself, i.e.,
\begin{equation}
d^a_C = \frac{\sum_{i=0}^{\sigma-1} \tilde{d}^{a-i}_C}{\sigma} \;
\end{equation}
for each time-step $a$ of the trajectory. This necessitates pruning the trajectory. The first $\sigma$ and last $\sigma$ time steps have to be included in the average, but cannot be included in the analysis itself, since they cannot be properly averaged. As such, larger $\sigma$ lead to slightly smaller trajectories being analyzed.

This smooth signal now allows for analyzing the equilibration behavior using these distances, with $d^a_C$ decreasing as more and more time steps are performed. Naturally, this analysis is build on the assumption that $\ubR_\mathrm{ref}$ can be considered equilibrated, but this can easily be confirmed by inspecting $g_\mathrm{ref}$. Furthermore, clear equilibration patterns become apparent when visualizing $d^a_C$, which helps in identifying unequilibrated trajectories. 

The next step in determining the first equilibrated configuration is to calculate a threshold for $d^a_C$. To do so, we assume a certain portion toward the end of the (pruned) trajectory to be equilibrated, as a fraction of the entire trajectory. Averaging $d^a_C$ across this portion $p_\mathrm{equi}=0...1$ of the distance metric, with usually $p_\mathrm{equi}$ around 0.1-0.2, yields an equilibration threshold $d_\mathrm{T}$. Once the distance metric falls below this threshold after $N_\mathrm{T}$ time steps, we assume the trajectory to be equilibrated henceforth.

\begin{figure}[h]
    \centering
    \includegraphics[width=0.9\columnwidth]{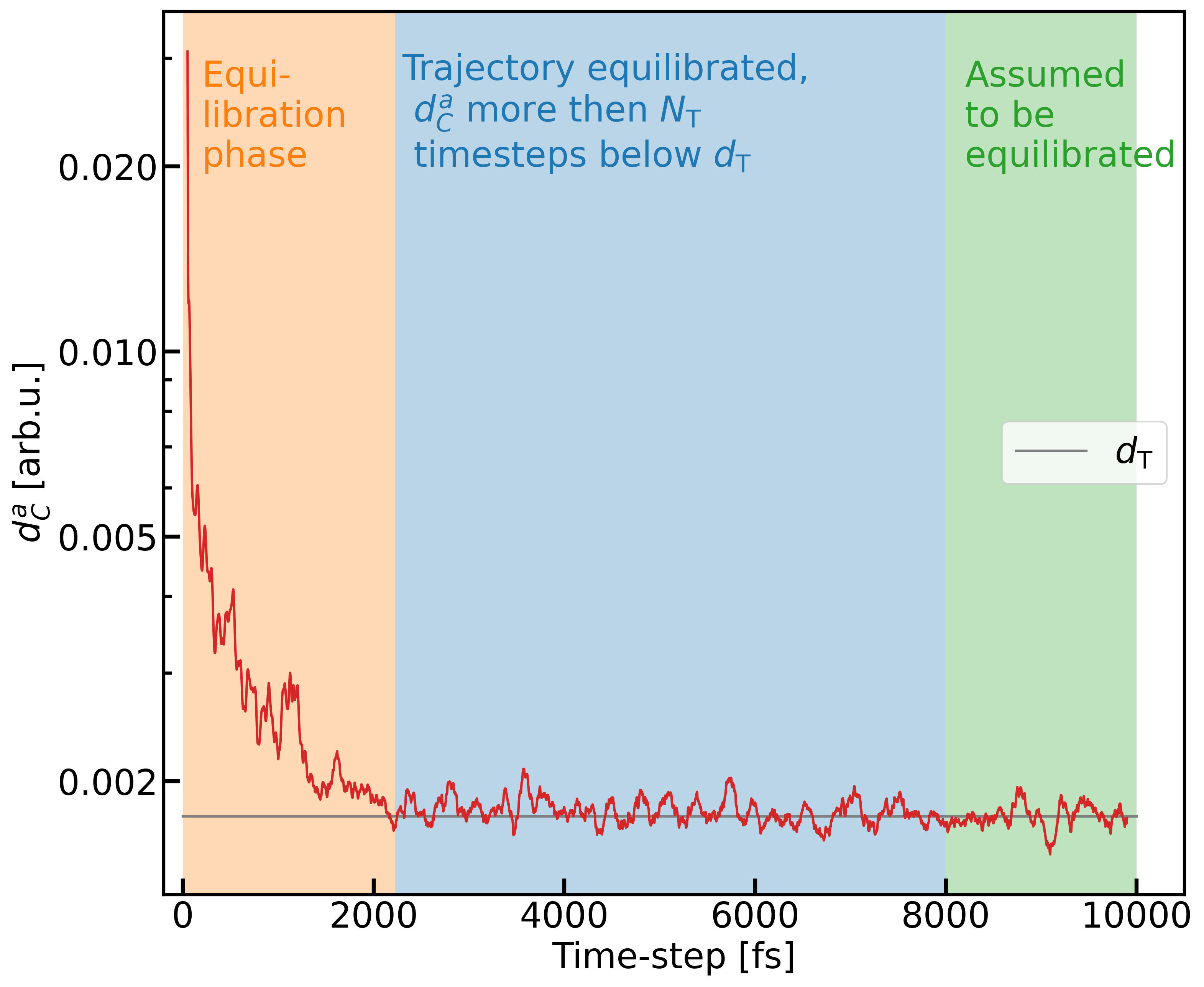}
    \caption{Convergence analysis employ throughout our work, shown for a trajectory of 512 Beryllium atoms at room temperature.}
    \label{fig:convergence_method}
\end{figure}

Calculating $d_\mathrm{T}$ from an averaged distance metric implies that a trajectory may be equilibrated slightly earlier then the algorithm will detect, since $d^a_C$ regularly exceeds $d_\mathrm{T}$ when fluctuating around the equilibrated average. Alternatively, one may set $d_\mathrm{T}$ according to the upper limit of such fluctuations (i.e., via the standard deviation), but this is problematic, as not fully equilibrated systems could be interpreted as such. We thus employ the average across the equilibrated configurations rather then the upper limit, to ensure that no unequilibrated system is misidentified. The first equilibrated configuration determined can therefore be interpreted as the point where a trajectory is equilibrated beyond doubt. The algorithm discussed here is further visualized in Fig.~\ref{fig:convergence_method}.

Please note that in its current form this technique is limited to analyzing fully equilibration patterns for trajectories which are fully equilibrated. Dynamical application of this method, i.e., determining whether a trajectory is equilibrated during run time, should be possible in principle, but require further development, exceeding the scope of this work. 

\subsection{DFT Surrogate models}
ML and data-driven methodologies are becoming increasingly important for DFT applications~\cite{deepdive}. Such approaches encompass interatomic potentials, which replace DFT as a means to evaluate the potential energy surface $A_\mathrm{tot}$ and forces $-\partial A_\mathrm{tot}/\partial \bR_\alpha$ in MD simulations, as well as property mappings, which learn specific physical and chemical properties from large DFT databases. We have recently~\cite{ellis_accelerating_2021} introduced an ML based workflow to reproduce DFT calculations in terms of both energy and electronic structure predictions, the Materials Learning Algorithms package (MALA)~\cite{mala}. MALA predicts the electronic structure of an ionic configuration via a local mapping that predicts the local density of states (LDOS) $d$ based on descriptors encoding local ionic configuration around points in real space. As such it is in principle capable of predicting the electronic structure of extended systems~\cite{size_transfer}. This prediction is realized by feed-forward neural networks (NNs).

Our current research is focused on applying such surrogate models to dynamical simulations or use them to perform thermodynamic sampling via Monte Carlo methods. However, it is well known that NNs perform poorly in extrapolation~\cite{extrapolatingNNs}, and should mostly be used for interpolation tasks. Therefore, optimal results require initialization of such calculations with configurations for which the model predictions will constitute an interpolation. This can be challenging when predicting the electronic structure of systems of extended size, for which no KS-DFT-MD data can be acquired, yet model predictions \textit{may} be interpolatory, given that physically sound ionic configurations are used.

\subsection{Hybrid MD workflow}
\begin{figure}[h]
    \centering
    \includegraphics[width=0.9\columnwidth]{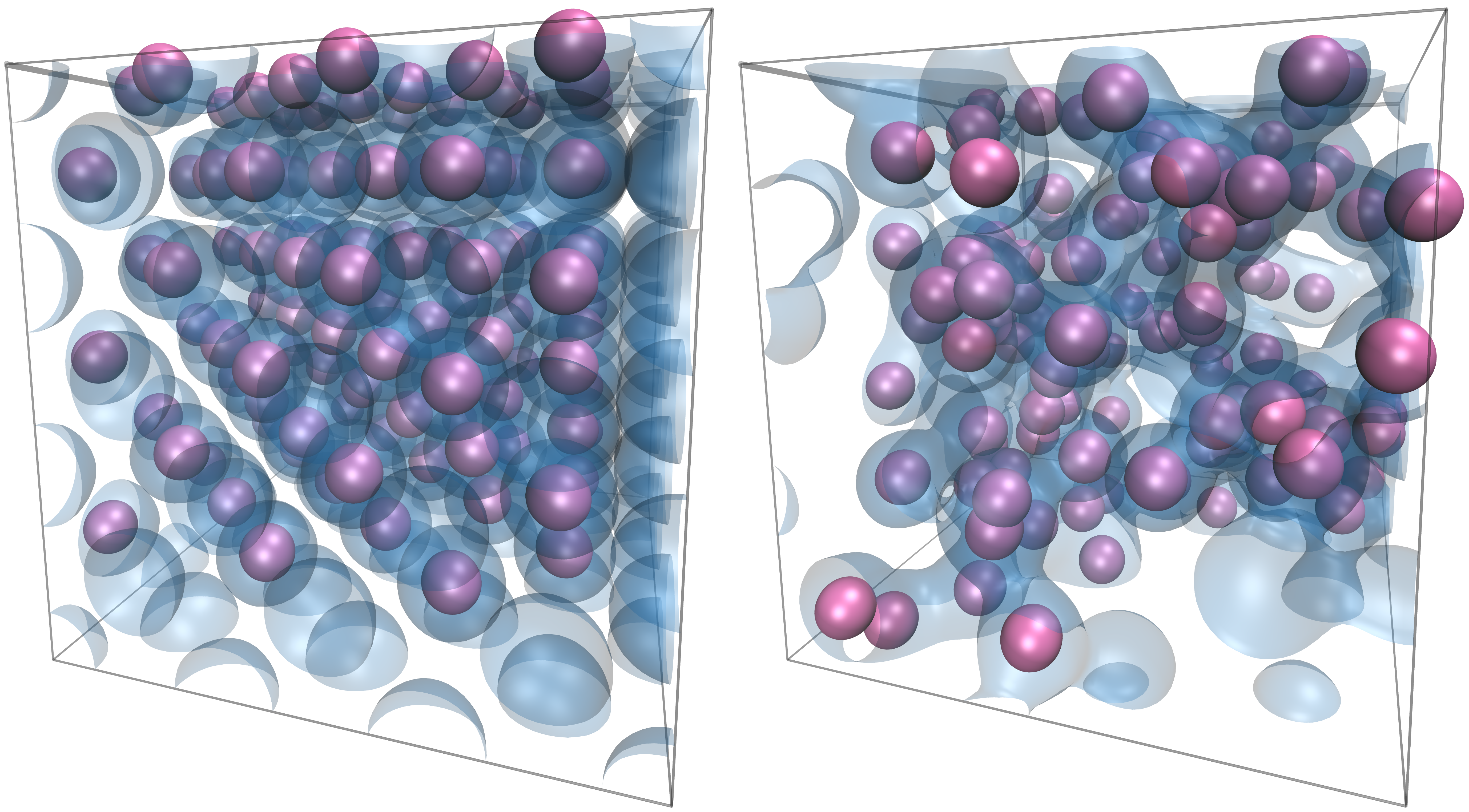}
    \caption{Ideal crystal structure (left) and equilibrated ionic configuration (right) of Hydrogen at $\tau=6.26$ eV, including an isosurface of the electronic density, created using VMD~\cite{VMD,Tachyon}.}
    \label{fig:densities}
\end{figure}

As the central methodological development, we present a hybrid OF-DFT-MD and KS-DFT-MD workflow for dynamical materials modelling. Usually, one has to rely on ideal crystal structures to initialize KS-DFT-MD trajectories or DFT surrogate model sampling routines. As we show in Sec.~\ref{sec:results}, OF-DFT-MD trajectories, which can be calculated at almost negligible computational cost and with favorable scaling behavior, serve as an alternative that improves performance in either case. Especially at larger temperatures, the actual ionic configurations and electronic densities differ greatly from ideal crystal structures, as shown in Fig.~\ref{fig:densities}. 

The task of OF-DFT-MD in our hybrid workflow is thus not to \textit{replace} KS-DFT-MD fully, as KS-DFT-MD trajectories are always built on top of the OF-DFT-MD generated structures. Rather, by providing reliable approximations of the actual ionic configurations, OF-DFT-MD (1) reduces the equilibration period for KS-DFT-MD trajectories and (2) improves surrogate model prediction accuracy. There is still some ``equilibration'' required to transition from slightly differing equilibrated configurations, but the associated number of time-steps and computational cost is vastly reduced. In practice, we simply choose the last configuration of the OF-DFT-MD trajectory as the initial configuration for KS-DFT-MD or DFT surrogate model evaluation.

\begin{figure*}[t!]
     \subfloat[Distance metrics for KS-DFT-MD and OF-DFT-MD, both starting from an ideal crystal structure. In either case, distance metrics are given relative to the equilibrated KS-DFT-MD configuration.\label{fig:ofdft_vs_ksdft}]{%
      \includegraphics[width=0.45\textwidth]{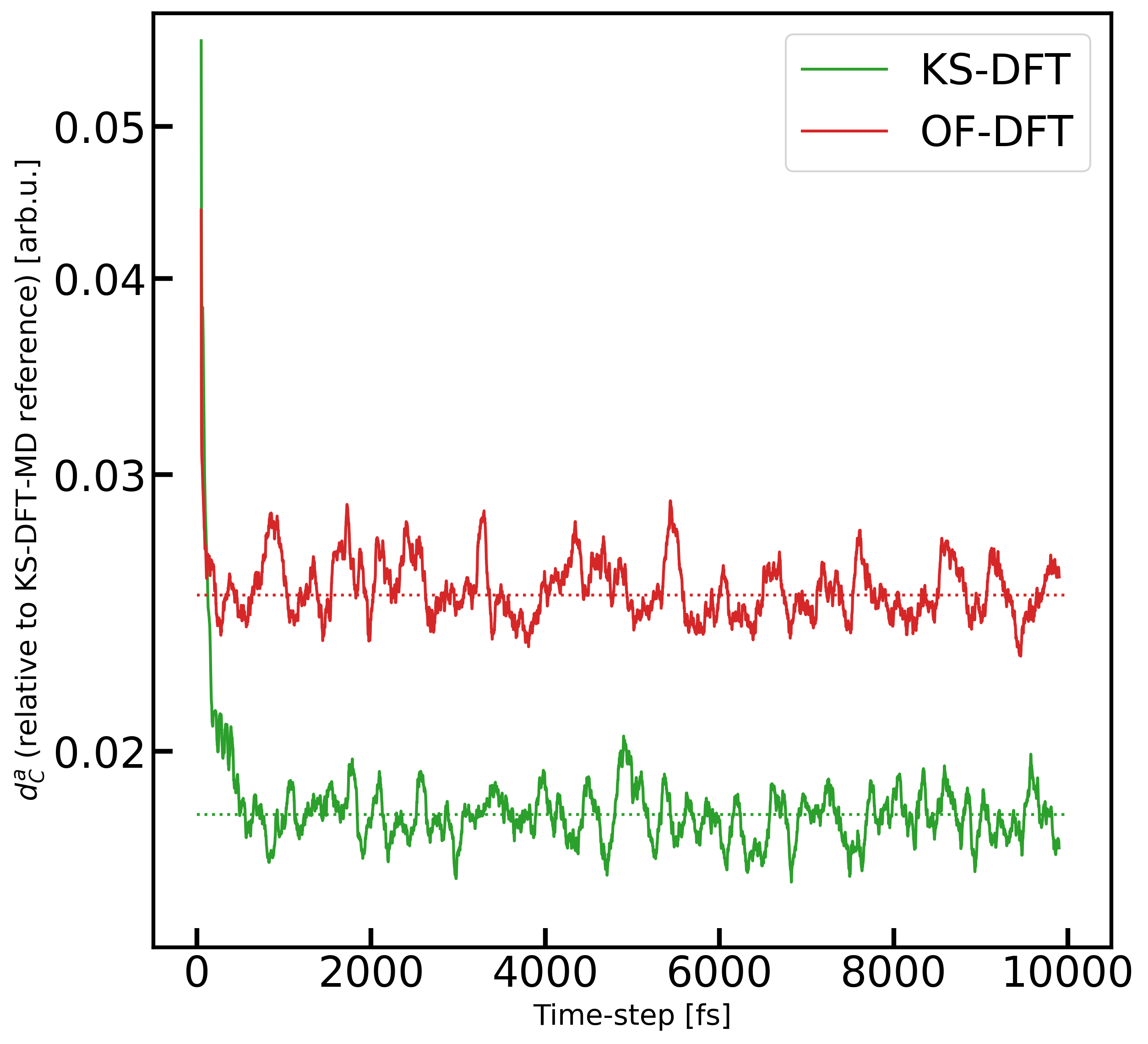}
    }
     \hfill
          \subfloat[Full visualization of the hybrid workflow approach. Blue: Equilibration from ideal crystal structure using OF-DFT-MD; Green: OF-DFT-MD propagation of equilibrated configuration; Red: Short ``equilibration'' phase in going from OF-DFT-MD to KS-DFT-MD. Orange: KS-DFT-MD sampling of configurations or thermodynamic observables.\label{fig:ofdft_plus_ksdft}]{%
      \includegraphics[width=0.45\textwidth]{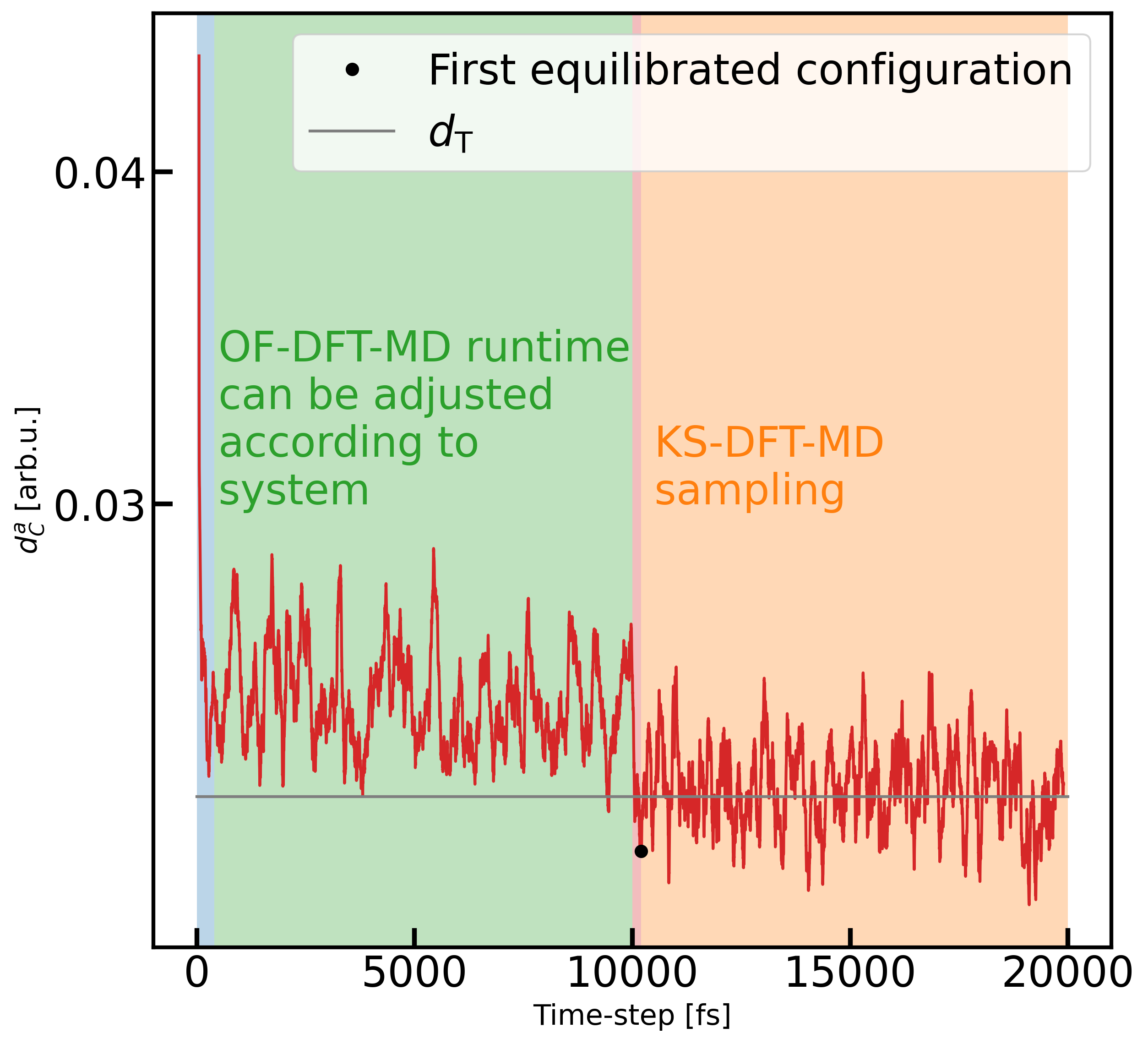}
    }
        \caption{Sketch of hybrid workflow based on the data for 128 Beryllium atoms. The data depicted here is the same as analyzed in Sec.~\ref{sec:length_scales}.}
        \label{fig:hybrid_workflow}
\end{figure*}

This workflow is further illustrated in Fig.~\ref{fig:hybrid_workflow}. There, Fig.~\ref{fig:ofdft_vs_ksdft} details the equilibration process in both KS-DFT-MD and OF-DFT-MD starting from the ideal crystal structure; both methods equilibrate to slightly different averages. However, the difference between the equilibrated ionic configurations is drastically smaller than the difference between equilibrated configuration and ideal crystal structure in either case, leading to the aforementioned speed up. In Fig.~\ref{fig:ofdft_plus_ksdft}, which visualizes the entirety of the hybrid workflow, this short ``equilibration'' phase in going from OF-DFT-MD to KS-DFT-MD is shown in a red background color. In our calculations, we have kept the number of OF-DFT-MD time-steps purposefully high to ensure that no unequilibrated structure enters the analysis. As a result, the OF-DFT-MD propagation of the equilibrated configurations shown in green is longer than necessary. In practice, it can be shortened significantly, because only the initial equilibration of the system (shown in blue) is relevant for the hybrid workflow. 

\subsection{Computational details}
\label{sec:compdetails}
All KS-DFT-MD simulations were performed using the highly efficient GPU implementation of VASP~\cite{kresse_ab_1993,kresse_efficiency_1996,kresse_efficient_1996, vaspgpu1, vaspgpu2} at the $\Gamma$-point. In all cases, a plane-wave basis set with a PAW pseudopotential~\cite{blochl_projector_1994,kresse_ultrasoft_1999} and the PBE exchange-correlation functional~\cite{perdew_accurate_1986,perdew_accurate_1992,perdew_generalized_1996} were employed. As cutoff energies for the plane wave basis set 440 eV, 248 eV, and 350 eV were used for Aluminum, Beryllium, and Hydrogen, respectively. For the single-point KS-DFT calculations in Sec.~\ref{sec:mlresults}, we employ Quantum ESPRESSO ~\cite{giannozzi_q_2020,giannozzi_quantum_2009,giannozzi_advanced_2017} with parameters consistent as described in  Ref.~\cite{ellis_accelerating_2021}. Likewise, the same ML model is used. All Beryllium and Aluminium calculations have been carried out at room temperature and ambient mass density, 1.896 g/cc and 2.699 g/cc, respectively, while the Hydrogen simulations have been performed at $\theta=[0.007,\;0.25,\; 0.5,\; 0.75]$, which equals $\tau=[0.977\,\mathrm{eV},\;3.132\,\mathrm{eV},\;6.264\,\mathrm{eV},\;9.390\,\mathrm{eV}]$, and a mass density corresponding to $r_s=2$. In the initialization of these calculations for Aluminum, Beryllium and Hydrogen, an fcc, hcp and fcc structure has been used, respectively. The number of orbitals in KS-DFT simulations are adjusted with the change in temperature to ensure the smallest occupation number is not higher then $10^{-6}$. 

If not otherwise noted, KS-DFT-MD simulations have been performed for 10,000 time-steps with either a time-step of 1 fs (Aluminum and Beryllium) or 0.01 fs (Hydrogen). For Beryllium, different values of $Q$ have been tested to confirm our findings across different thermostat settings, while for Hydrogen, $Q$ has been kept fixed at 0.5. For Aluminum, $Q=0.001$ has been used. Beryllium trajectories have been analyzed with the aforementioned method and an assumed equilibrated portion of the trajectory $p_\mathrm{equi}=0.2$, a running average width $\sigma=100$, and required number of time-steps below the distance threshold for equilibration $N_\mathrm{T}=50$. As the higher temperature in the Hydrogen simulations lead to noisier signals, $N_\mathrm{T}$ and $\sigma$ were both increased to 200. We provide full equilibration graphs in App.~\ref{sec:app_curves_Beryllium} and App. ~\ref{sec:app_curves_Hydrogen} to further verify the choice of these parameters.

For Aluminum and Beryllium, the Wang-Teter kinetic energy functional~\cite{PhysRevB.45.13196} with PBE as exchange-correlation functional was used. Optimized effective potentials~\cite{localpsps} were used as local pseudopotentials. Given the emphasis on higher temperatures, for Hydrogen we used the finite temperature TF free energy functional with the von Weizsäcker gradient correction in combination with LDA XC functional. In order to perform finite-temperature OF-DFT-MD simulations, we implemented the finite-temperature TF functional in the OF-DFT code DFTpy, which was previously introduced for the ground state calculations~\cite{DFTpy}.
The version of DFTpy with the newly added option for the finite-temperature TF free energy functional is now available with the most recent open-source version of DFTpy. The underlying theoretical  details of our implementation are outlined in App.~\ref{sec:FT-TF}.
In all OF-DFT-MD runs, the MD time step was $2~{\rm fs}$ and all other simulation parameters were kept consistent with those used for the KS-DFT-MD calculations.

All DFT-MD data, example input scripts, and processing scripts can be obtained from Ref.~\cite{fiedler_dataset_2022}. All ML experiments and equilibration analysis have been carried out with the MALA code version 1.1.0~\cite{mala}.

\begin{figure*}[ht]
    \centering
    \includegraphics[width=0.9\textwidth]{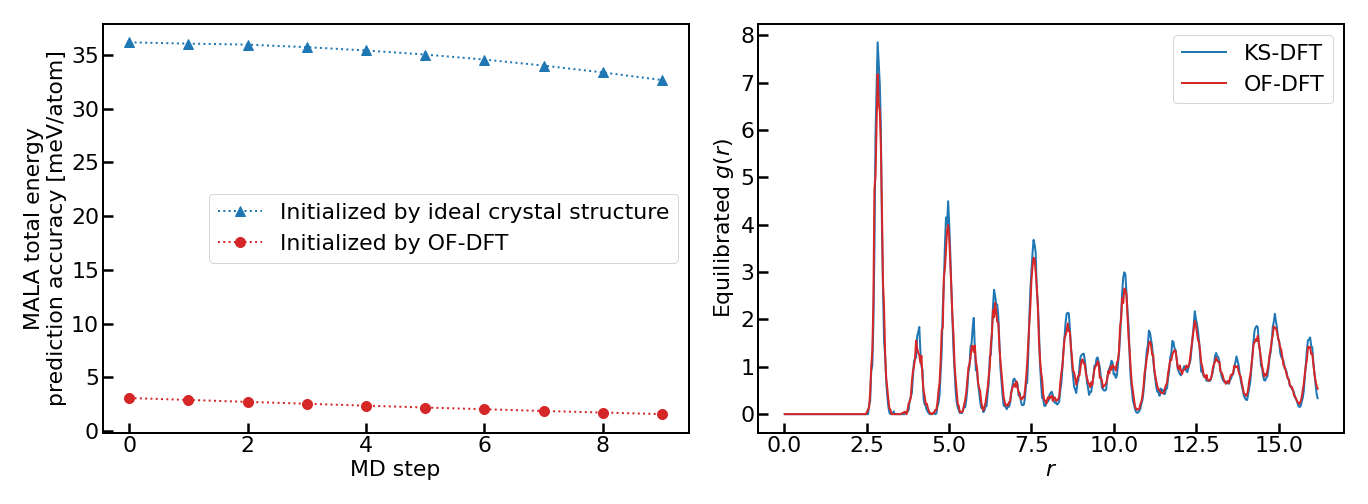}
    \caption{Left: Accuracy of ML surrogate models predictions on KS-DFT-MD trajectories that are initialized with the ideal crystal geometry and OF-DFT-MD. Right: Comparison of the RDF of equilibrated ionic configurations for 256 Aluminum atoms obtained from OF-DFT-MD and from KS-DFT-MD starting from the ideal crystal structure (RDF calculated for one configuration per method).}
    \label{fig:ml_init}
\end{figure*}

\section{Results}
\label{sec:results}
The results are divided into three parts, reflecting three areas of materials modeling for which our workflow is highly useful: (1) the application of ML methods trained on DFT data, (2) simulations with large numbers of atoms with KS-DFT-MD, and (3) simulations of systems at high temperatures using KS-DFT-MD. The utlity of the latter two applications is twofold, as improved KS-DFT-MD performance does not only alleviate the cost of direct sampling of thermodynamic obervables, but further helps with data generation for future ML applications.

\begin{figure*}[t!]
    \centering
    \includegraphics[width=0.9\textwidth]{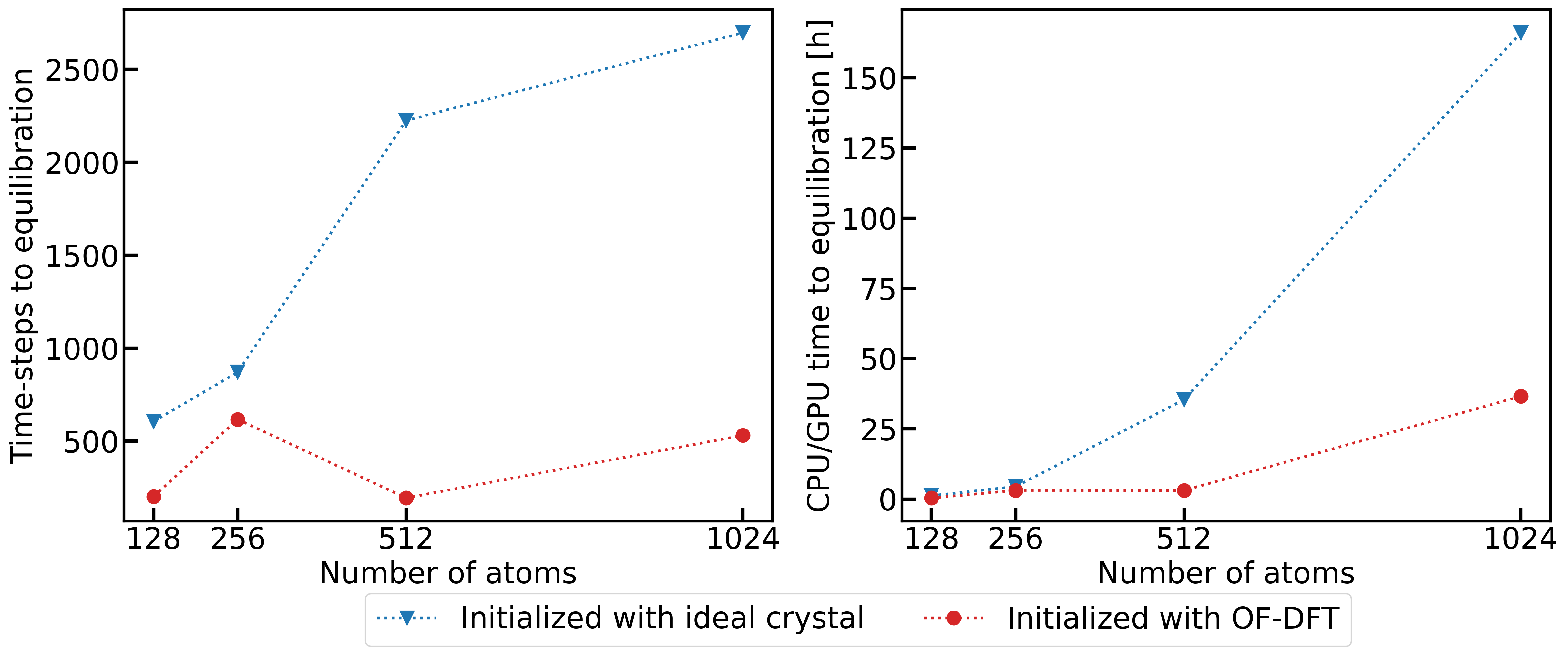}
    \caption{Equilibration of Beryllium simulation cells of increasing size according to our distance metric analysis.}
    \label{fig:init_results}
\end{figure*}

\subsection{ML-DFT trajectory initialization}
\label{sec:mlresults}
As mentioned above, one potential pitfall of transferable ML surrogate models for DFT calculations is their application to ionic configurations that are too dissimilar from observed training data for the model to perform well. Quantifying uncertainties in model predictions is naturally a crucial tool for detecting such behavior, but it does not solve the problem of providing initial configurations on which the model can be expected to perform reasonably well. One may incorporate ideal crystal structure data as well as data from equilibration phases into the training set to alleviate this problem -- but ultimately the interest lies with equilibrated configurations, and as such, one would unnecessarily complicate model training. Any such model would have to perform well on ideal crystals, slowly equilibrating systems as well as the actually equilibrated system. While such models are by no means unachievable (in fact, interatomic potentials often follow this approach), the amount of additional training data can become challenging if the actual electronic structure has to be learned rather than the potential energy surface. We thus propose to circumvent the need to do so by employing ionic configurations from OF-DFT-MD as initial configurations for extended DFT surrogate model driven simulations. Given that OF-DFT-MD captures the KS-DFT geometries on which the ML model has been trained, one can rely on ML predictions to equilibrate the system fully. We have investigated this behavior for Aluminum at room temperature. The results are shown in Fig.~\ref{fig:ml_init}, where we have used our MALA framework introduced above as a representative ML surrogate model. Two different KS-DFT-MD trajectories, each for 10 time steps, are investigated. One is initialized by OF-DFT-MD and the other by the ideal crystal structure. For each configuration in these trajectories, a full KS-DFT calculation is performed alongside a MALA inference (using the 298 K model from Ref.~\cite{ellis_accelerating_2021}) to determine the total free energies. Additional KS-DFT calculations are necessary, since the model was trained on data generated with Quantum ESPRESSO, while we perform the MD simulation with VASP to be consistent with the other results throughout this investigation.

As shown in Fig.~\ref{fig:ml_init}, within the first ten time steps, the configurations (and therefore prediction accuracies) per trajectory change only slightly. This means that for the ideal crystal, we have configurations consistent with 0 K, while for the OF-DFT-MD trajectory, we have configurations consistent with 298 K, with the model having only been trained on 298 K data. It can be seen that this temperature difference amounts to an error of around 35 meV/atom in the total free energy for the ideal crystal trajectory, as the configurations at 0 K represent an out-of-distribution sampling for the ML model, while the OF-DFT-MD initialized trajectory leads to in-distribution sampling. It is well known that NN based approaches excel at the latter task, while performing poorly at the former. In either case, the error reduces as the KS-DFT-MD trajectory equilibrates. When OF-DFT-MD is used to initialize the ionic configuration, the error is below 5 meV/atom from the beginning, which is consistent the errors reported in Ref.~\cite{ellis_accelerating_2021} for a full KS-DFT-MD trajectory. 

Naturally, the merit of this workflow mainly depends on the ability of OF-DFT-MD to converge to useful and comparable geometries. One way to visually confirm this is in terms of the RDF. As shown in Fig.~\ref{fig:ml_init}, both KS-DFT-MD and OF-DFT-MD generally agree quite well. While the amplitude of certain peaks is under- or overestimated by the latter, the overall positions match, which is the basis for the performance observed in Fig.~\ref{fig:ml_init}. As shown in the following sections, this behavior can be supported for increasingly larger systems and temperatures, two important dimensions in which one seeks to build transferable ML models. 

Overall, using OF-DFT-MD allows us to create models based solely on equilibrated data and apply them on extended scales using approximately correct initial configurations, thus saving model training time while maintaining high prediction accuracy.

\begin{figure*}[t!]
    \centering
    \includegraphics[width=0.9\textwidth]{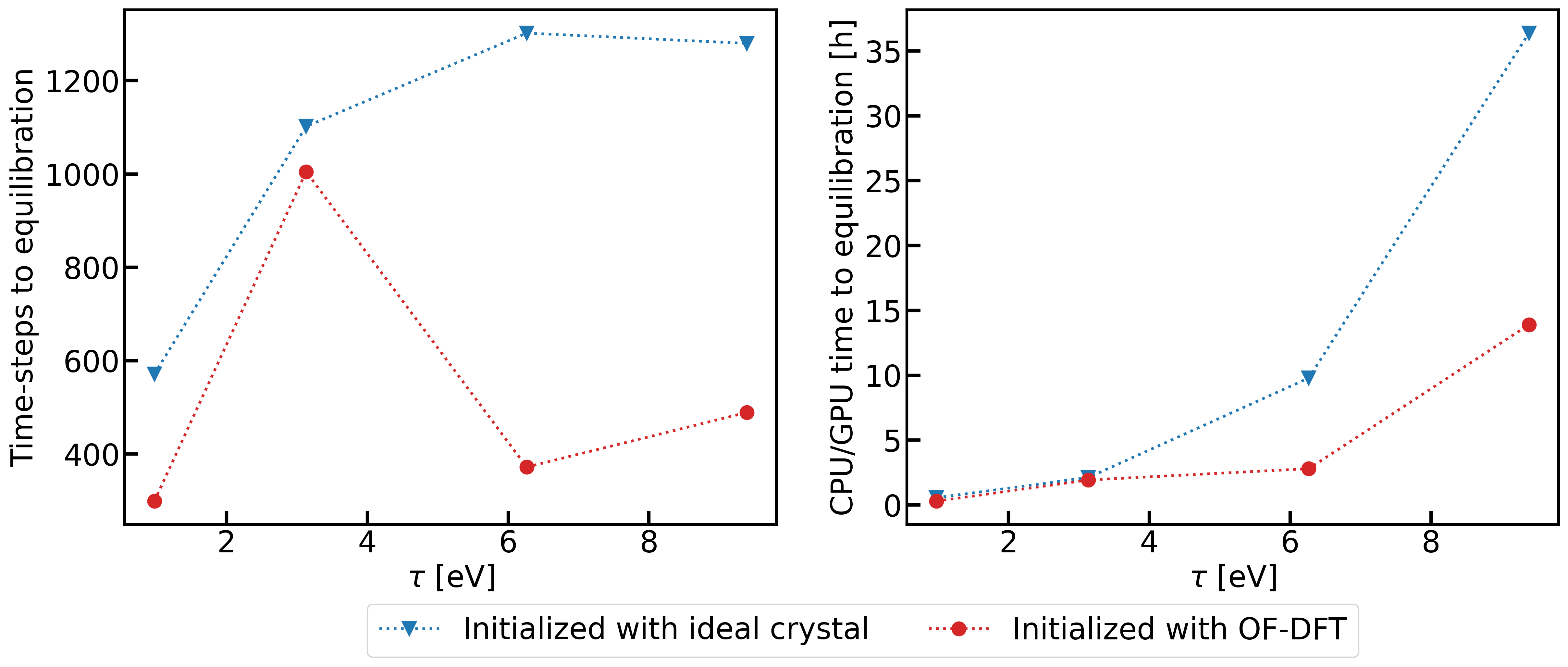}
    \caption{Equilibration of Hydrogen simulations for increasing temperature according to our distance metric analysis.}
    \label{fig:temp_results}
\end{figure*}

\subsection{Large length scales}
\label{sec:length_scales}
Numerical calculations with KS-DFT-MD are prone to finite-size errors~\cite{finiteSizeErrors}, if small simulation cells are used. Due to the aforementioned scaling behavior of DFT, increasingly larger simulations are computationally challenging. This is especially problematic given the fact that for large simulation cells a larger number of time steps is usually required to properly equilibrate the system, yet each time-step is more costly by itself. Decreasing the time to equilibration would therefore drastically improve the performance of such simulations, and can be achieved with OF-DFT-MD. To test our workflow in this setting, we equilibrate Beryllium simulation cells of increasing size at room temperature and ambient mass density starting from the ideal crystal structure and OF-DFT-MD trajectories. For each trajectory we determine the time required to equilibration using the algorithm outlined in Sec.~\ref{sec:equianalysis}. The results of this analysis are given in Fig.~\ref{fig:init_results}. Generally, OF-DFT-MD yields an almost perfectly equilibrated geometry. Equilibration is detected after a number of time-steps almost of the same magnitude as $N_\mathrm{T}$, i.e., the minimum resolution our analysis can provide. Our analysis thus unveils no significant overall equilibration to be necessary for the OF-DFT-MD initialized trajectories, as is exemplified by the full equilibration curves provided in App.~\ref{sec:app_curves_Beryllium}. Furthermore, while a clear increase in initialization time can be seen for trajectories initialized with ideal crystal structure, no such trend is observed for those initialized with OF-DFT-MD. As DFT calculations become increasingly costly with system size, this grows to large amounts of computational time being saved. In the case of 1024 Beryllium atoms, we are able to save more then 100 GPU hours, which correspond to approximately 20\% of the entire simulation time. The increase in time-steps required for equilibration is linear with the number of particles in the case of ideal crystal initialization.

Equilibration depends on the thermostat used during the DFT-MD run. As we use the Nosé-Hoover thermostat, the equilibration time is chiefly influenced by the choice of the Nosé mass $Q$. We have verified that our findings are not impacted by the choice of this technical parameter, see App.~\ref{sec:app_nose_mass}

These speed ups are offset by the time needed to perform OF-DFT-MD simulations. These simulation however accrue to only a small computational overhead compared to KS-DFT-MD simulations. For example, a single time-step costs $t\simeq 3.8~{\rm s}$ ($t\simeq 13~{\rm s}$) for 128 (1024) Beryllium atoms using a single CPU, or about 11 (36) hours using one CPU without parallelization for the entire trajectory of 10,000 time steps. While these times may seem excessive by itself, they are easily explained by two important considerations. Firstly, as can be seen above for the case of KS-DFT-MD, 10,000 times-steps are not required for equilibrated configurations under these conditions, we have chosen such a number to ensure that our investigations are not distorted by unequilibrated trajectories. Furthermore, we have obtained all OF-DFT-MD results using a single CPU; such timings do not compare to GPU hours reported for KS-DFT-MD trajectories, even when reporting core-hours rather then wall-time hours. We thus omit OF-DFT-MD timings from Fig.~\ref{fig:init_results} and \ref{fig:init_results_full}.

\subsection{High temperatures}
\label{sec:temperatures}
As discussed above, KS-DFT is known to scale unfavorably with temperature. To investigate our hybrid workflow for increasingly larger temperatures, we thus choose a computationally more tractable system then Beryllium, namely 108 Hydrogen atoms under a range of temperatures up to $\theta=0.75$, i.e., $\tau=9.39\,\mathrm{eV}$. We then undertake a similar investigation as in the preceding section, by performing KS-DFT-MD simulations at increasing temperatures, initialized either by the ideal crystal structure or OF-DFT-MD ionic configurations, and analyzing the equilibration patterns. The results are illustrated in Fig.~\ref{fig:temp_results}.

The temperature dependence of the increase in time steps required to equilibrate a system is not as straightforward as the dependence on the number of particles reported in Fig.~\ref{fig:init_results}. Larger temperature fluctuations lead to noisier trajectories, making equilibration, and determination of achieving it, more difficult. The necessity to use smaller simulation cells at the large temperatures investigated further leads to additional noise. This noise is witnessed in the equilibration curves App.~\ref{sec:app_curves_Hydrogen}. Fig.~\ref{fig:temp_results} shows a slight temperature-dependence of the number of time-steps required to equilibrate a trajectory from an ideal crystal structure. For the OF-DFT-MD initialized trajectory, such a dependence cannot be clearly deduced. Rather, we observe that as temperature increases, the number of time-steps saved by the utilizing an OF-DFT-MD initialized trajectory appears to be roughly constant (with the exception of the result for $\tau=6.264$ eV, which seems to be an outlier). As the computational cost per individual time step increases with temperature, OF-DFT-MD initialization leads to sizable savings in computation time. As the influence of different thermostat parameters has amply been investigated in App.~\ref{sec:app_nose_mass}, we restrict ourselves to only one thermostat parameter here. 

In the case of Hydrogen, one OF-DFT-MD step for 108 particles requires about $t \simeq 2.5~{\rm s}$ regardless of the temperature value. This results in approximately 7 hours on a single CPU for 10,000 time-steps. Apart from this, the same considerations as discussed in Sec.~\ref{sec:length_scales} hold true, i.e., less time steps might be necessary in actuality. 

Since OF-DFT-MD calculation times do not increase with temperature, while KS-DFT scales so unfavorably due to increase in the number of recurred bands (orbitals), OF-DFT-MD initializations can be crucial when attempting to investigate larger temperature ranges, i.e., in the WDM regime. We see already from the results in Fig.~\ref{fig:temp_results}, that the saved computational time increases massively for moving to $\tau=9.39\,\mathrm{eV}$, for which the main cause is the drastically increased cost per time-step as temperature increases.

\section{Conclusion and Outlook}\label{sec:end}
We have shown that initializing ionic configurations based on  OF-DFT-MD greatly reduce the computational cost for equilibrating KS-DFT-MD trajectories. Given that OF-DFT is in principle able to capture the physics of the systems at hand, KS-DFT-MD thereafter requires only very little computational overhead to reach equilibration. We have verified our results on different systems and have shown that our findings hold true as both the system size and temperature increase. Our hybrid workflow is thus widely applicable to simulations of matter under extreme conditions, especially given the favorable scaling properties of OF-DFT-MD with temperature and system size.

OF-DFT-MD is an especially crucial addition to the toolkit of first-principles simulations when it comes to ML-DFT applications. Not only does it improve performance of data acquisition, as outlined in Sec.~\ref{sec:length_scales} and \ref{sec:temperatures}, but furthermore it directly helps with ensuring that models are use in an interpolative manner, by providing initial configurations on which surrogate model inference yields to accurate predictions. 

Yet, there is some work to be done. OF-DFT-MD workflows have to be integrated into larger frameworks, such as MALA. It is further well known that OF-DFT is not universal in the types of systems that can be treated. While metals can generally be treated to high accuracy~\cite{DFTpy}, other systems evade accurate treatment. Theoretical development and the advent of ML based kinetic energy functionals may alleviate these problems. 
Our presented hybrid approach, where accuracy of the OF-DFT part is essential for the acceleration of the simulations (but not for the accuracy of the final KS-DFT results), represents another motivation for further developing kinetic energy and entropy functionals for OF-DFT applicable to various materials.  

\section*{Acknowledgments}
This work was partially supported by the Center for Advanced Systems Understanding (CASUS) which is financed by Germany’s Federal Ministry of Education and Research (BMBF) and by the Saxon state government out of the State budget approved by the Saxon State Parliament.
\appendix

\section{Finite-Temperature Thomas-Fermi Free Energy}
\label{sec:FT-TF}
The TF approximation is particularly useful at high temperature. We therefore provide the theoretical basis of the finite-temperature TF free energy functional, which we have implemented in the OF-DFT code DFTpy~\cite{DFTpy}.

In general, the noninteracting free energy and the free-energy density are defined as
\begin{equation}\label{T_0}
F_{s}[n]\equiv T\s[n] - \kB\tau S\s[n]=\int d\br f_s[n;\tau](\br).
\end{equation} 
In TF theory, we have $f_s \equiv f_{\rm TF}$, where $f_{\rm TF}$ is the Thomas-Fermi free energy density~\cite{PhysRevE.91.033104, PhysRevA.20.586} defined as
\begin{equation}\label{TF}
f_{\rm TF}\left([n],\theta\right)
= \frac{\sqrt{2}m^{3/2}}{\hbar^3\pi^2\beta^{5/2}}\left(\eta I_{1/2}(\eta)-\frac{2}{3}I_{3/2}(\eta) \right),
\end{equation} 
where $I_{\nu}$ is the Fermi integral of order $\nu$, and  $\eta=\mu \beta $ is a constant following from the normalization  $N=\int n(\vec r) \mathrm{d}\vec{r}$ and $\beta=1/(k_B\tau)$. 
The Fermi-Dirac integral of order $\nu$ is defined as
\begin{equation}
I_{\nu}=\int_0^{\infty} {\mathrm{d}x}~\frac{x^{\nu}}{(1+\exp(x-\eta))}.
\end{equation}
The functional derivative of the free energy with respect to the electron density yields~\cite{POP2018}
\begin{equation}\label{TF_pres}
v_s[n(\vec r); \tau]=\frac{\delta F_{\rm TF}}{\delta n}=\mu [n(\vec r); \tau].
\end{equation} 
\begin{figure}
    \centering
    \includegraphics[width=0.45\textwidth]{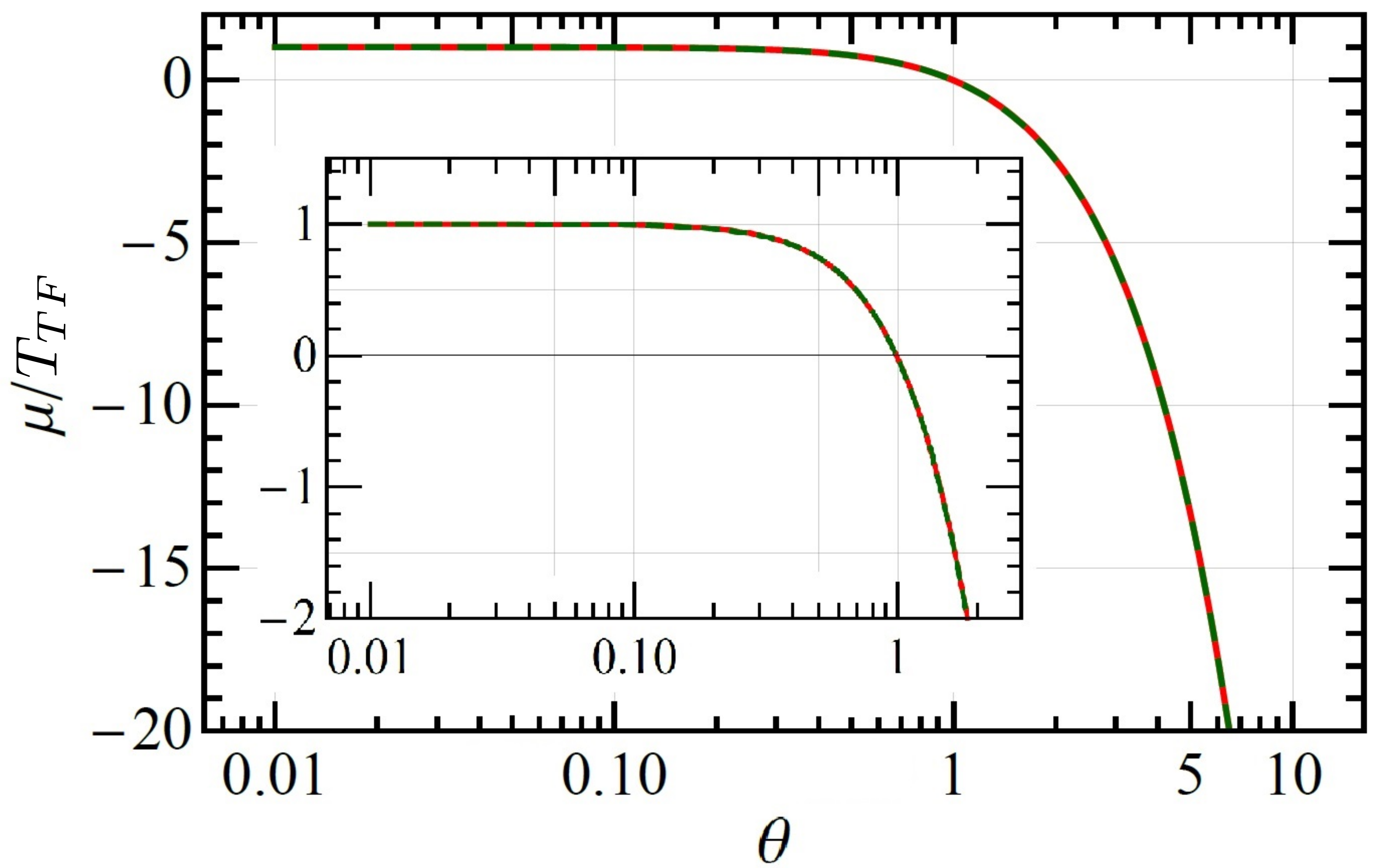}
    \caption{Chemical potential inversion according to the numerical solution of the inverse problem in Eq.~(\ref{eq:eta}) (red) and obtained from the approximation in Eq.~(\ref{eq:fit}) (green).}
    \label{fig:chem}
\end{figure}
In order to evaluate the effective potential in Eq.~(\ref{TF_pres}), we need to find the chemical potential of the free, noninteracting electron gas at finite temperature which is formally given by the inverse of~\cite{PhysRevA.29.1471}
\begin{equation}\label{eq:eta}
    \frac{2}{3} \left(\theta[n(\vec r)]\right)^{-3/2}= \int_0^{\infty} {\mathrm{d}x}~\frac{\sqrt{x}}{(1+\exp(x-\eta))}.
\end{equation}
From Eq.~(\ref{eq:eta}), we see that $\eta[n]= \beta \mu [n]$ is defined by the degeneracy parameter given by $\theta[n]=k_B\tau/T_{TF}[n]$. 
To compute $\mu  [n(\vec r);\tau]$, one can use the following expression (in the Hartree atomic units) as the solution of the inverse problem Eq.~(\ref{eq:eta})~\cite{Sotnikov}:
\begin{widetext}
\begin{equation}\label{eq:fit}
\mu  [n(\vec r);\tau] = \begin{cases}
\alpha n(\vec r)^{2/3}\left(1+\sum_{i}^4 a_i\times \left(\frac{\tau}{\alpha n(\vec r)^{2/3}}\right)^i\right) &\text{if $\theta= \frac{\tau}{\alpha n(\vec r)^{2/3}}< 1.36$}\\
\tau~\ln \left( C \left(\frac{\tau}{\alpha n(\vec r)^{2/3}}\right)^{-3/2}\right)+\tau\ln \left( 1+C \left(2\frac{\tau}{\alpha n(\vec r)^{2/3}}\right)^{-3/2}\right)~ &\text{if $\theta=\frac{\tau}{\alpha n(\vec r)^{2/3}}\geq 1.36$}
\end{cases}
\end{equation}
\end{widetext}
where $\alpha= \frac{(3\pi^2)^{2/3}}{2}$,
$a_1 = 0.016$, $a_2 = -0.957$, $a_3 = -0.293$, $a_4 = 0.209$, and $C=\frac{2}{3 \Gamma(3/2)}=0.752252778063675$.

Note that the the dimensionless potential $v_s/(\alpha n(\vec r)^{2/3})=\mu /(\alpha n(\vec r)^{2/3})$ depends only on the parameter $\theta [n(\vec r);\tau] =\frac{\tau}{\alpha n(\vec r)^{2/3}}$, the local degeneracy parameter. This dependence is shown in Fig. (\ref{fig:chem}), where we
compare the exact solution of the inverse problem Eq.(\ref{eq:eta}) for $\mu$ with the approximation in Eq.~(\ref{eq:fit}). This comparison demonstrates the high accuracy of the approximation in Eq.~(\ref{eq:fit}). 

We also observe in Fig. \ref{fig:chem}, that in the limit of low temperatures $\theta=\frac{\tau}{\alpha n(\vec r)^{2/3}}\ll 1$, Eq. (\ref{eq:fit}) follows the TF model
\begin{equation}
    v_s= \mu[n(\vec r)]=T_{TF}[n]=\alpha n(\vec r)^{2/3}.
\end{equation}
At high temperatures, $\theta>1$, the TF potential yields negative values. This can lead to numerical instabilities in the OF-DFT scheme, where the total free energy is minimized. This problem is solved by introducing a gradient correction which is strictly positive and blocks the emergence of large density gradients.

For the simulation of Hydrogen at high temperatures, we used the von Weizsäcker functional defined in Eq. (\ref{eq:vW}) as a gradient correction.
One can use different versions of a functional representing first order gradient correction to the local density approximation (e.g., see Refs~\cite{Jones_1971, PhysRevA.20.586, Kirzhnits_1975, POP2015, POP2018}). However, at the considered high temperatures, both the ion-electron and ion-ion coupling are weak and the ion-ion interaction is not so sensitive to the corrections beyond the LDA~\cite{PhysRevE.98.023207, PhysRevE.99.053203, cpp_22_zm, cpp_17_zm}. In this case, the role of a gradient correction with OF-DFT merely amounts as a numerical trick for stabilizing the numerics.

In the simulations involving non-zero temperatures the von Weizs{\"a}cker and nonlocal terms are borrowed from the corresponding zero temperature expressions.

\section{Equilibration curves for Beryllium}
\label{sec:app_curves_Beryllium}
\begin{figure*}[ht]
    \centering
    \includegraphics[width=0.9\textwidth]{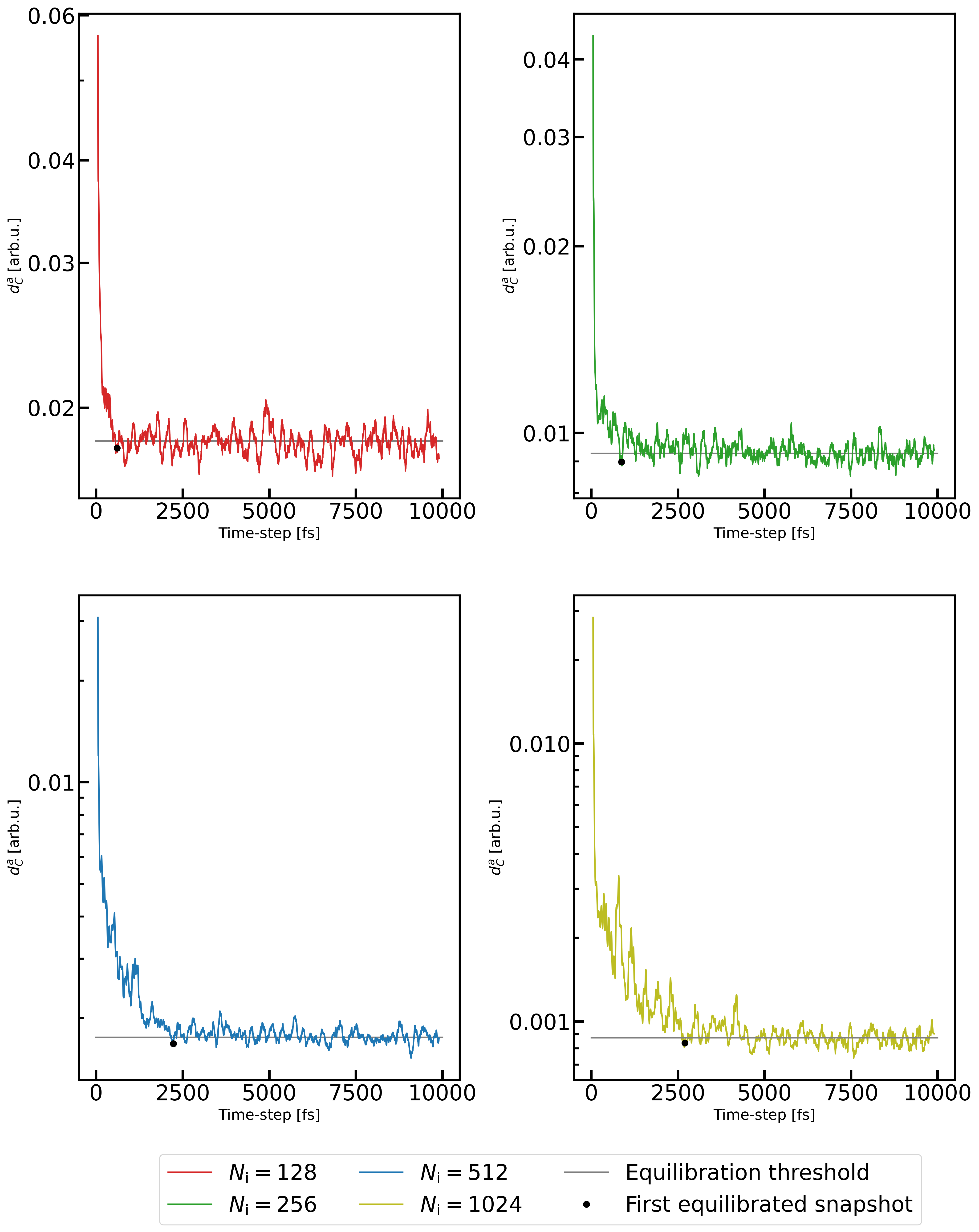}
    \caption{Equilibration curves for Beryllium, starting from the ideal crystal structure.}
    \label{fig:equi_be_ideal}
\end{figure*}

\begin{figure*}[ht]
    \centering
    \includegraphics[width=0.9\textwidth]{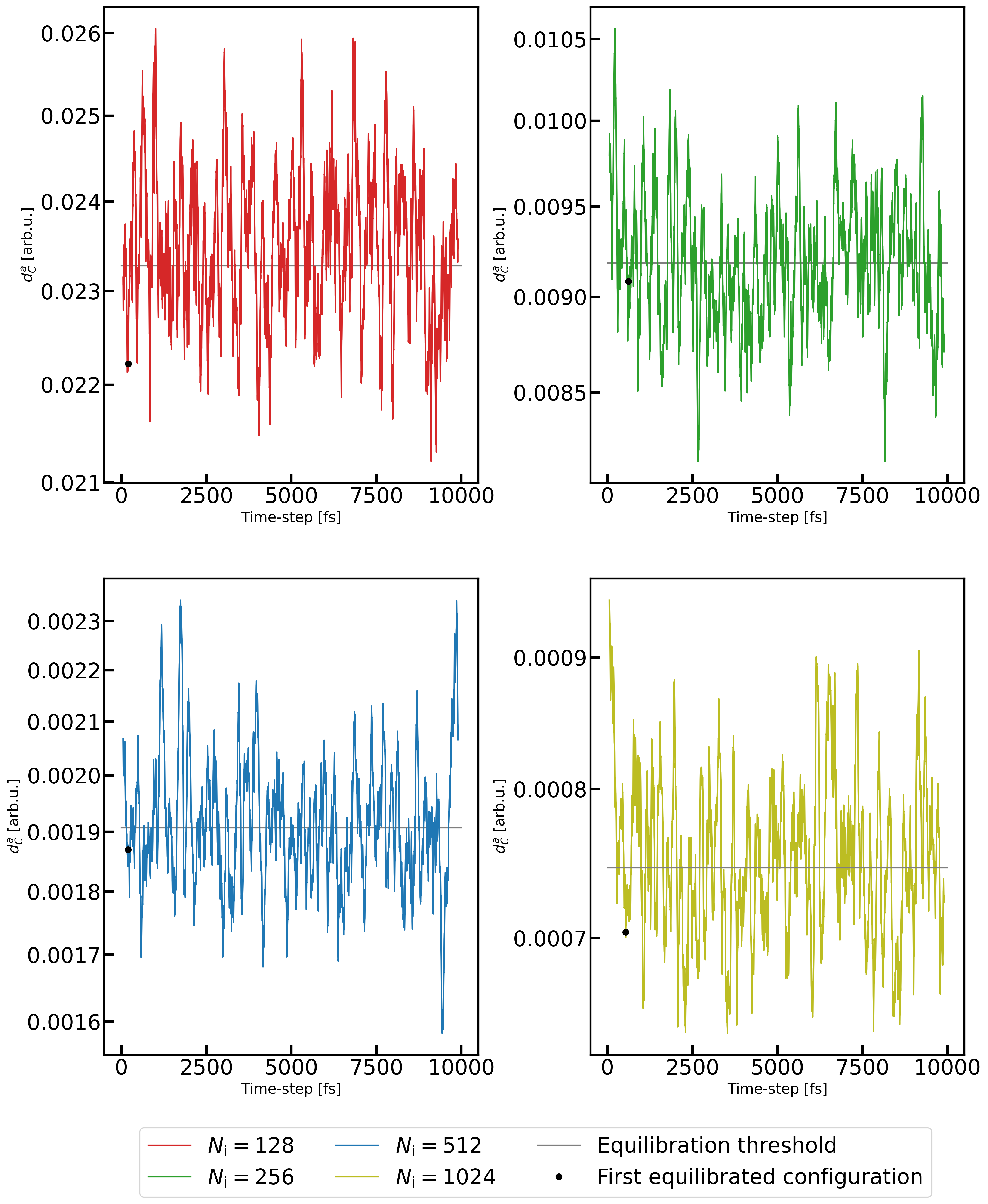}
    \caption{Equilibration curves for Beryllium, starting from the OF-DFT-MD structure.}
    \label{fig:equi_be_ofdft}
\end{figure*}

\begin{figure*}[ht]
    \centering
    \includegraphics[width=0.9\textwidth]{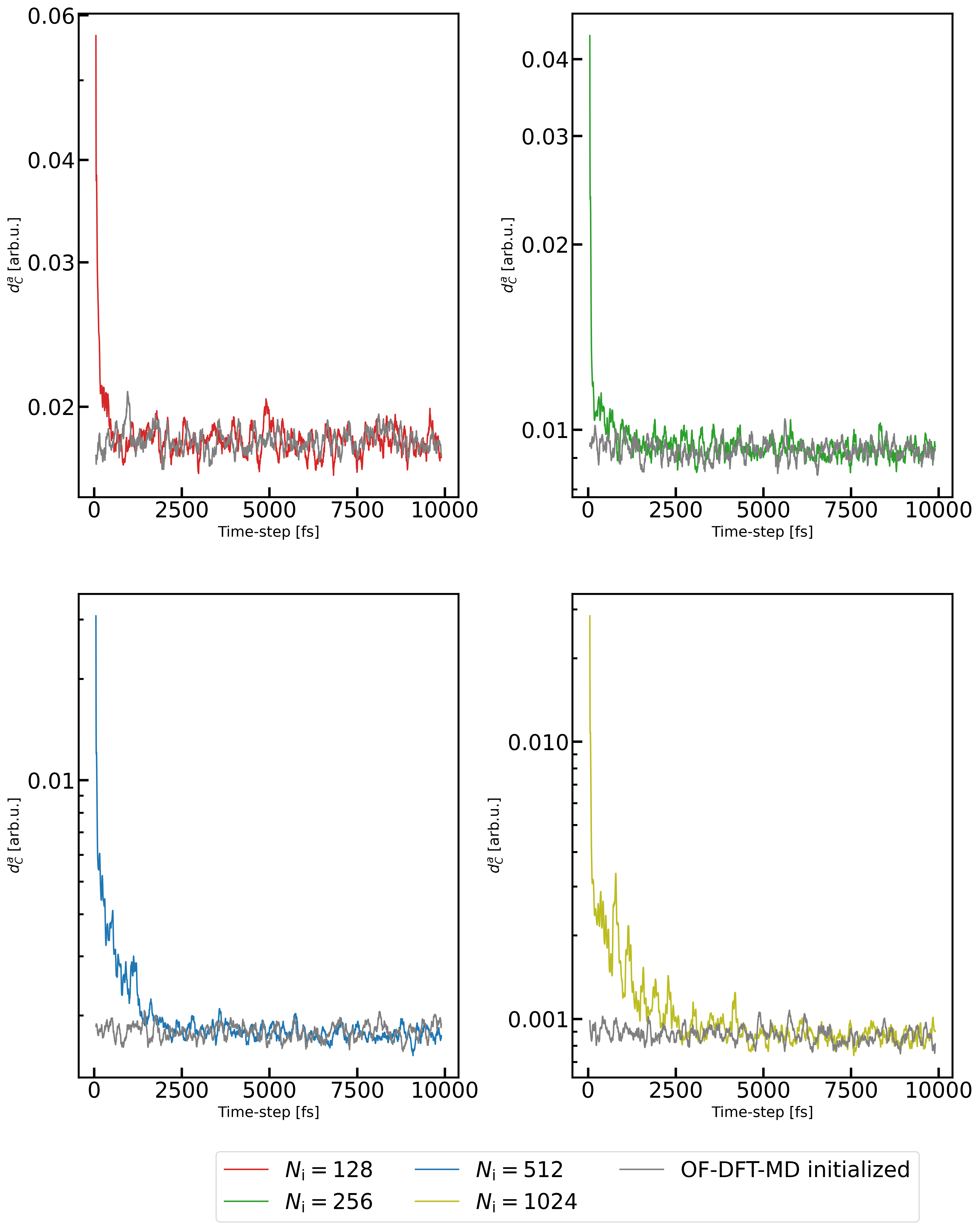}
    \caption{Equilibration curves for Beryllium, with the OF-DFT-MD initiliazed trajectories superimposed over those starting from ideal crystal structure. Please note that to be consistent, both distance metrics were calculated with the reference configuration of the ideal crystal structure trajectory, and thus, the OF-DFT-MD initialized metrics differ slightly form those shown in Fig.~\ref{fig:equi_be_ofdft}.}
    \label{fig:equi_be_relative}
\end{figure*}

In the following, we provide the full equilibration curves for the Beryllium based experiments presented above, to confirm the correct application of the algorithm outlined in Sec.~\ref{sec:equianalysis} and Fig.~\ref{fig:convergence_method}. The given figures follow Fig.~\ref{fig:convergence_method} in their general outline, i.e.~the smoothed cosine distance between configurations and reference configurations is shown, along with equilibration thresholds and first equilibrated configuration. The equilibration behavior for Beryllium is shown in Fig.~\ref{fig:equi_be_ideal} for DFT-MD simulations starting from the ideal crystal structure and in Fig.~\ref{fig:equi_be_ofdft} for those started from OF-DFT-MD configurations. In Fig.~\ref{fig:equi_be_ideal} the gradual equilibration process is neatly pronounced, it can be seen how the system is slowly thermalized, and how thermalization requires more and more time as one moves to larger temperatures. Please note that raw numeric values of $d^a_C$ should not be compared between individual systems, as they are relative quantities. Absolute values of these metrics differ across system sizes since the small deviations in the RDFs become increasingly small relative to the overall RDF. This however does not negatively affect performance, as only relative values are compared. Equilibration trends are notably less pronounced in Fig.~\ref{fig:equi_be_ideal} as the initial configurations are already very close to the equilibrated ones. A slight equilibration can be seen for 1024 atoms, where the first few hundred time-steps are mostly above the equilibration threshold. Generally however, the OF-DFT-MD trajectories are very close to equilibrium to begin with, which is reflected in the similar (and rather small) number of time-steps required for equilibration throughout. The distance metrics essentially fluctuate around the average almost from the beginning of the simulation. Fig.~\ref{fig:equi_be_relative} corroborates that both methods of initialization lead to equilibrated trajectories that fluctuate around the same average. 

\section{Equilibration curves for Hydrogen}
\label{sec:app_curves_Hydrogen}
\begin{figure*}[ht]
    \centering
    \includegraphics[width=0.9\textwidth]{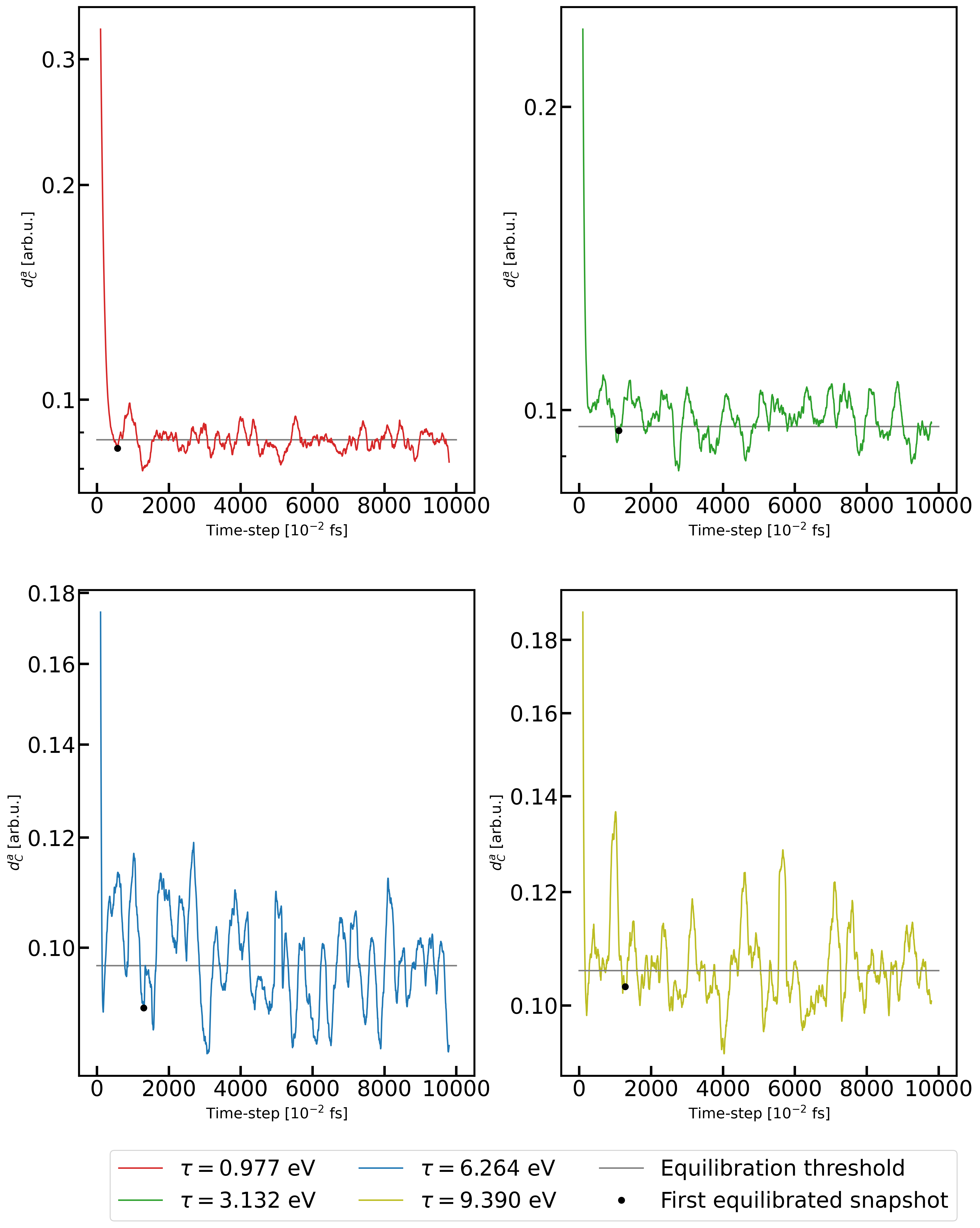}
    \caption{Equilibration curves for Hydrogen, starting from the ideal crystal structure.}
    \label{fig:equi_h_ideal}
\end{figure*}

\begin{figure*}[ht]
    \centering
    \includegraphics[width=0.9\textwidth]{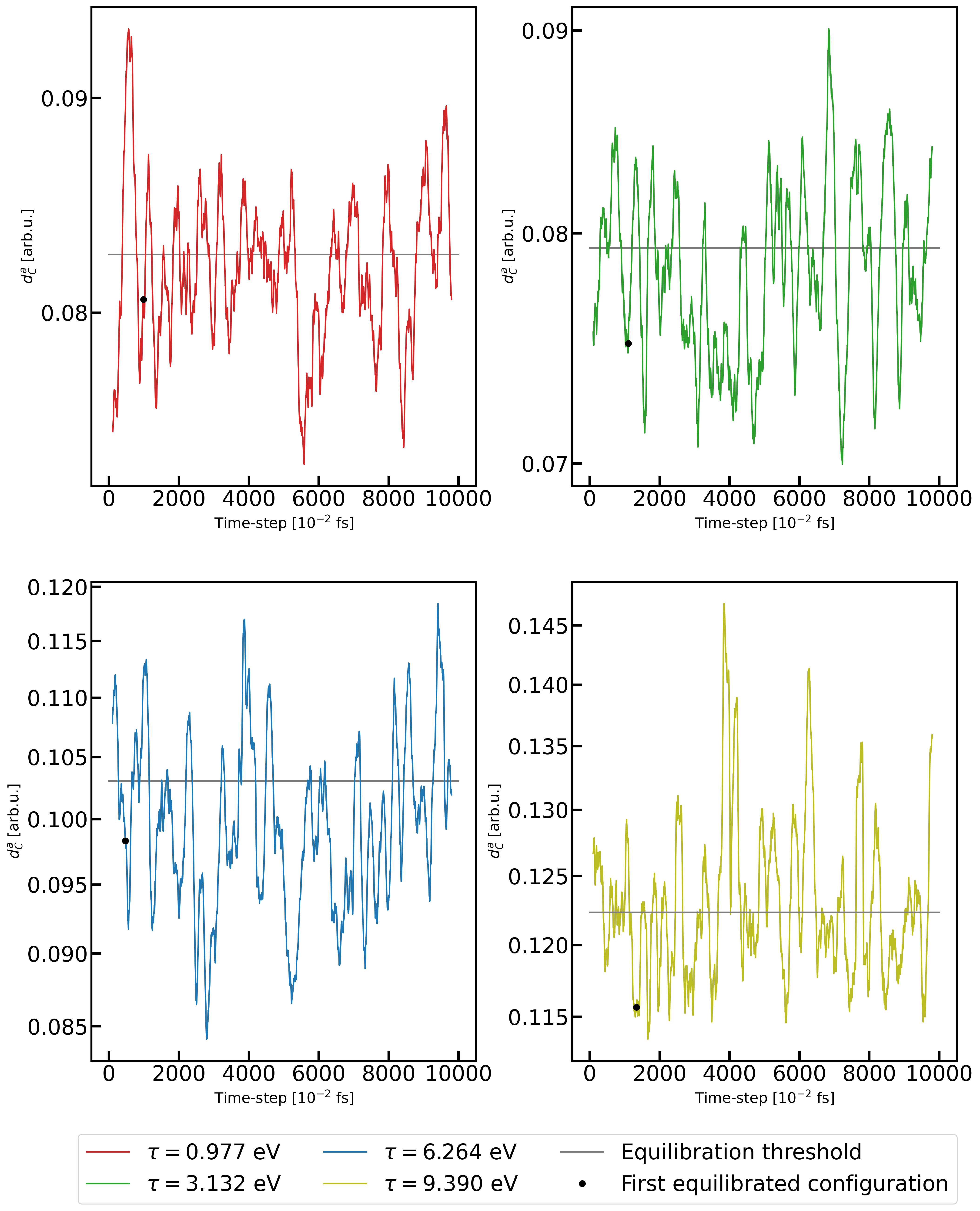}
    \caption{Equilibration curves for Hydrogen, starting from the OF-DFT-MD structure.}
    \label{fig:equi_h_ofdft}
\end{figure*}

\begin{figure*}[ht]
    \centering
    \includegraphics[width=0.9\textwidth]{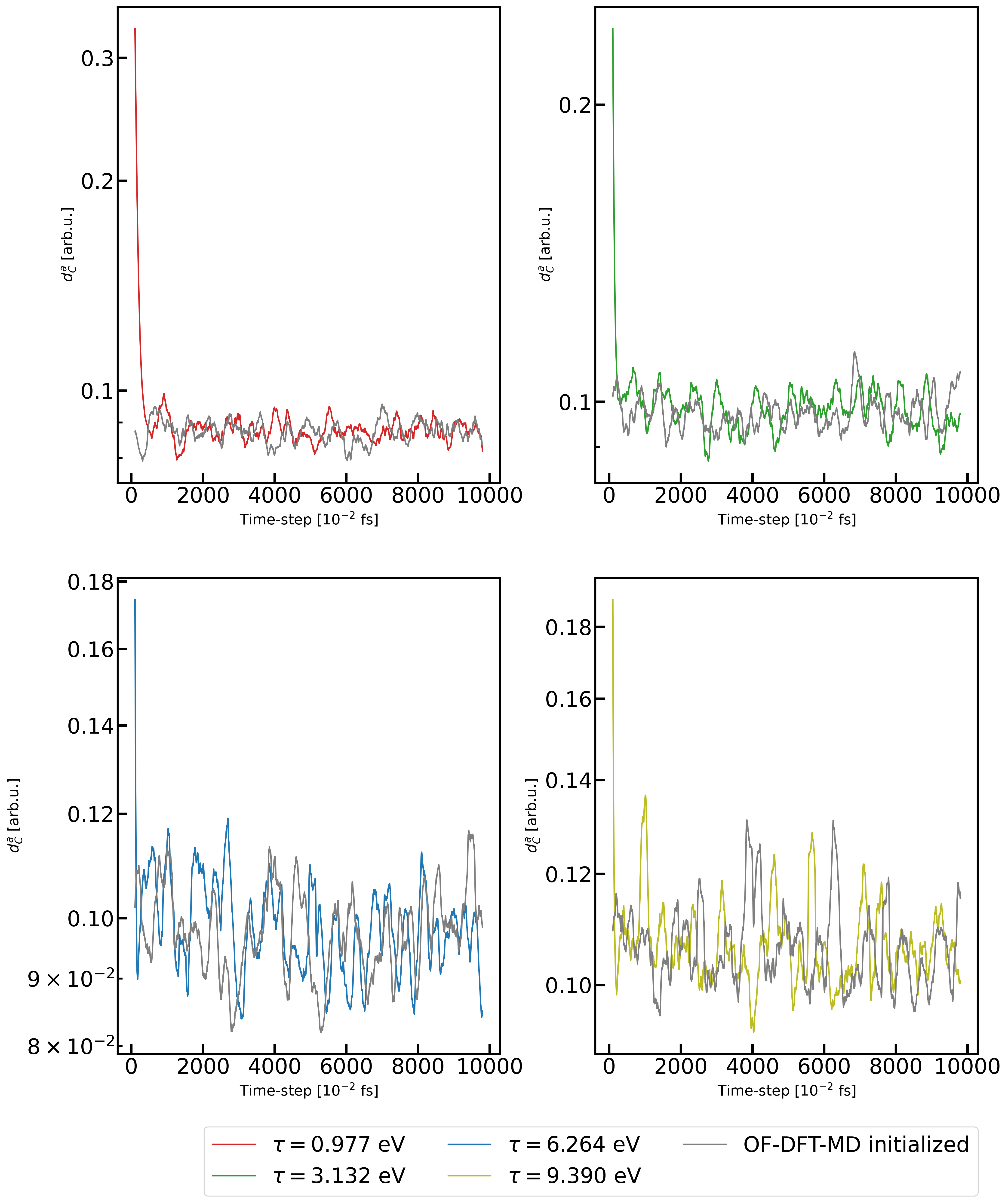}
    \caption{Equilibration curves for Hydrogen, with the OF-DFT-MD initiliazed trajectories superimposed over those starting from ideal crystal structure. Please note that to be consistent, both distance metrics were calculated with the reference configuration of the ideal crystal structure trajectory, and thus, the OF-DFT-MD initialized metrics differ slightly form those shown in Fig.~\ref{fig:equi_h_ofdft}.}
    \label{fig:equi_h_relative}
\end{figure*}

Similar to App.~\ref{sec:app_curves_Beryllium}, here we provide the full equilibration curves for the Hydrogen based experiments presented in Sec.~\ref{sec:temperatures}. The equilibration behavior for Hydrogen is shown in Fig.~\ref{fig:equi_h_ideal} for DFT-MD simulations starting from the ideal crystal structure and in Fig.~\ref{fig:equi_h_ofdft} for those starting from OF-DFT-MD simulations. The overall equilibration behavior is similar to those observed in the preceding section. In contrast to the results for Beryllium, it is noticeable that the trajectories appear to be noisier, even with additional smoothing, obfuscating the equilibration process to a degree, especially visible in Fig.~\ref{fig:equi_h_ofdft}. While this in principle \textit{can} lead to a slight overestimation of the overall equilibration phase, it does not infringe on our analysis, as the same parameters are employed to analyze both the ideal crystal and OF-DFT-MD initialized trajectories. Furthermore we minimize such effects by increasing $N_\mathrm{T}$ and $\sigma$ for larger $\tau$. Generally, the Hydrogen trajectories equilibrate faster, which is mostly due to the smaller number of atoms. Fig.~\ref{fig:equi_h_relative} shows that either method of initialization converges to the same fluctuating average. 

\section{Influence of Nosé mass}
\label{sec:app_nose_mass}

\begin{figure}[h]
    \centering
    \includegraphics[width=0.9\columnwidth]{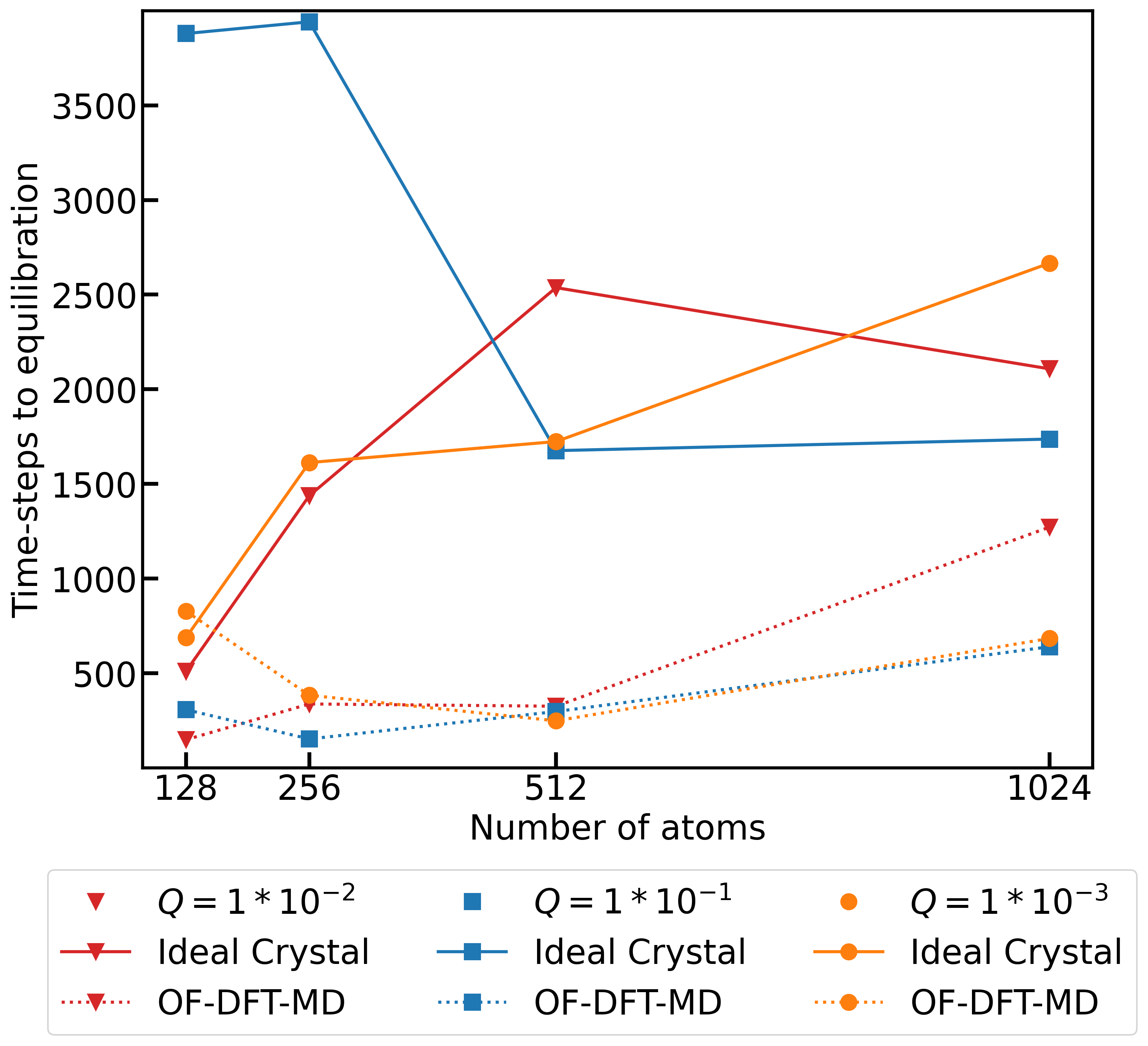}
    \caption{Comparison of MD runs with different $Q$. In contrast to Fig~\ref{fig:init_results}, all MD trajectories were run for 5000 time-steps of 1 fs for computational feasibility. Please note that the less systematic behavior of the $Q=0.1$ and $Q=0.005$ trajectories can be explained by the trajectories not being fully converged after 5000 time-steps, and that the results for $Q=0.01$ vary slightly compared to Fig.~\ref{fig:init_results}, since in Fig.~\ref{fig:init_results} the full 10000 time-steps were analyzed. Here only 5000 time-steps are analyzed, in order to be consistent with the other trajectories.}
    \label{fig:init_results_full}
\end{figure}

As the performance of DFT-MD simulations that employ the Nosé-Hoover thermostat depend on $Q$ we have to verify the applicability of the results presented in Fig.~\ref{fig:init_results} across the range of applicable $Q$. More precisely, for Beryllium, we determine the outer boundaries of values for $Q$ for which the KS-DFT-MD trajectories do not diverge (i.e., do not find equilibrium in the NVT ensemble or produce temperature oscillations) as $Q=[0.1, 0.005]$. Comparing to the boundaries of this interval, $Q=0.01$, which has been used for the simulations shown in Fig.~\ref{fig:init_results}, gives reasonably good performance for the equilibration. This comparison is shown in Fig.~\ref{fig:init_results_full}, and illustrates that our choice of $Q=0.01$ reflects standard production quality. It further shows that OF-DFT-MD reduces computational time needed to equilibrate a system across different values of $Q$. This trend can be expected to hold true for other thermostats as well. 

\clearpage
\newpage

%

\end{document}